\documentclass[sigconf]{acmart}

\pdfoutput=1

\usepackage[utf8]{inputenc} 
\usepackage[T1]{fontenc}    
\usepackage{hyperref}       
\usepackage{url}            
\usepackage{booktabs}       
\usepackage{amsfonts}       
\usepackage{nicefrac}       
\usepackage{microtype}      
\usepackage{xcolor}         

\usepackage{enumitem}
\usepackage{xcolor}
\usepackage{tikz}
\usepackage{pgfplots}
\usepackage[multiple]{footmisc}
\usepackage{multirow}
\usepackage{bm}
\usepackage{booktabs}
\usepackage{array}
\usepackage{subcaption}
\usepackage{graphicx}
\usepackage{tabularx}
\usepackage{lipsum}
\usepackage{algorithm}
\usepackage{algorithmicx}
\usepackage{algpseudocode}
\usepackage{soul}
\usepackage{float}
\usepackage{tcolorbox}
\usepackage[flushleft]{threeparttable}
\usepackage{tablefootnote}
\usetikzlibrary{pgfplots.groupplots}

\usepackage{tabularx}
\usepackage{geometry}
\usepackage{lipsum}
\usepackage{pifont}
\usepackage{pdfrender}

\newcommand{\cmark}{\ding{51}}%

\pgfplotsset{
    name nodes near coords/.style={
        every node near coord/.append style={
            name=#1-\coordindex,
            alias=#1-last,
        },
    },
    name nodes near coords/.default=coordnode
}

\pgfplotsset{compat=1.5.1}
\def\addlegendimage{\csname pgfplots@addlegendimage\endcsname}
\usetikzlibrary{patterns}
\usetikzlibrary{pgfplots.groupplots}
\usetikzlibrary{
  pgfplots.colorbrewer,
}

\definecolor{aa}{rgb}{0.2,0.7,0.310}
\definecolor{cc}{rgb}{1.0,0.49,0.0}
\definecolor{bb}{rgb}{0.514,0.325,0.831}

\newcommand{\aka}{{\it a.k.a.}}
\newcommand{\ie}{{\textit i.e.}}
\newcommand{\eg}{{\textit e.g.}}
\newcommand{\ours}{\texttt{FairGAD}}
\newcommand{\dominant}{\textsc{DOMINANT}}
\newcommand{\conad}{\textsc{CONAD}}
\newcommand{\cola}{\textsc{CoLA}}
\newcommand{\vgod}{\textsc{VGOD}}
\newcommand{\fairod}{\textsc{FairOD}}
\newcommand{\correlation}{\textsc{Correlation}}
\newcommand{\hin}{\textsc{HIN}}
\newcommand{\fairwalk}{\textsc{FairWalk}}
\newcommand{\edits}{\textsc{EDITS}}

\newcommand\red[1]{\textcolor{red}{#1}}
\newcommand\blue[1]{\textcolor{blue}{#1}}
\newcommand\violet[1]{\textcolor{violet}{#1}}

\AtBeginDocument{%
  }

\copyrightyear{2024}
\acmYear{2024}
\setcopyright{acmlicensed}\acmConference[CIKM '24]{Proceedings of the 33rd
ACM International Conference on Information and Knowledge
Management}{October 21--25, 2024}{Boise, ID, USA}
\acmBooktitle{Proceedings of the 33rd ACM International Conference on
Information and Knowledge Management (CIKM '24), October 21--25, 2024,
Boise, ID, USA}
\acmDOI{10.1145/3627673.3679754}
\acmISBN{979-8-4007-0436-9/24/10}

\title{Towards Fair Graph Anomaly Detection: \\ Problem, Benchmark Datasets, and Evaluation}

\author{Neng Kai Nigel Neo}
\authornote{Both authors contributed equally to this research.}
\affiliation{%
  \institution{Georgia Institute of Technology}
  \city{Atlanta, Georgia}
  \country{USA}
}
\email{nnnk@gatech.edu}

\author{Yeon-Chang Lee}
\authornotemark[1]
\affiliation{%
  \institution{Ulsan National Institute of Science and Technology (UNIST)}
  \city{Ulsan}
  	\country{Korea}
}
\email{yeonchang@unist.ac.kr}

\author{Yiqiao Jin}
\affiliation{%
  \institution{Georgia Institute of Technology}
  \city{Atlanta, Georgia}
  \country{USA}
}
\email{yjin328@gatech.edu}

\author{Sang-Wook Kim}
\affiliation{%
 \institution{Hanyang University}
 \city{Seoul}
  	\country{Korea}
 }
\email{wook@hanyang.ac.kr}

\author{Srijan Kumar}
\authornote{Corresponding author.}
\affiliation{%
  \institution{Georgia Institute of Technology}
  \city{Atlanta, Georgia}
  \country{USA}
}
\email{srijan@gatech.edu}

\renewcommand{\shortauthors}{Neng Kai Nigel Neo, Yeon-Chang Lee, Yiqiao Jin, Sang-Wook Kim \& Srijan Kumar}

\begin{document}

\begin{abstract}
The Fair Graph Anomaly Detection (\ours) problem aims to accurately detect anomalous nodes in an input graph while avoiding biased predictions against individuals from sensitive subgroups. 
However, the current literature does not comprehensively discuss this problem, nor does it provide realistic datasets that encompass actual graph structures, anomaly labels, and sensitive attributes.
To bridge this gap, we introduce a formal definition of the \ours\ problem and present two novel datasets constructed from the social media platforms Reddit and Twitter. These datasets comprise 1.2 million and 400,000 edges associated with 9,000 and 47,000 nodes, respectively, and leverage political leanings as sensitive attributes and  misinformation spreaders as anomaly labels.
We demonstrate that our \ours\ datasets significantly differ from the synthetic datasets used by the research community. 
Using our datasets, we investigate the performance-fairness trade-off in nine existing GAD and non-graph AD methods on five
state-of-the-art 
fairness methods.
Code and datasets are available at \href{https://github.com/nigelnnk/FairGAD}{https://github.com/nigelnnk/FairGAD}. 
\end{abstract}



\maketitle

\section{Introduction}\label{sec:introduction}

\begin{table*}[t]
\small
\centering
\caption{Statistics of relevant datasets and our \ours\ datasets. 
Note that the synthetic graph structure was constructed based on edges formed by structural similarities between nodes (see~\citet{agarwalUnifiedFrameworkFair2021} for the details).
That is, our \ours\ datasets are new comprehensive benchmark datasets that cover all of \violet{graph}, \blue{anomaly detection}, and \red{fairness} aspects in the real world.
}
\vspace{-0.2cm}
\label{tab:statistics}
\setlength{\tabcolsep}{1.5pt} 
\resizebox{0.95\textwidth}{!} {
 \renewcommand{\arraystretch}{1.1}
\begin{tabular}{@{}c|cccccc|ccc|ccc||cc@{}}
\toprule
\multicolumn{1}{c|}{\multirow{2}{*}{\textbf{Dataset}}}                 & \multicolumn{6}{c|}{\textbf{GAD}~\citep{liu2022bond}} & \multicolumn{3}{c|}{\textbf{Fairness}~\citep{DaiW21,fair_edits}} & \multicolumn{3}{c||}{\textbf{Fair Non-graph AD}~\citep{agarwalUnifiedFrameworkFair2021}}      & \multicolumn{2}{c}{\textbf{FairGAD}} \\ 
        & \textbf{Weibo} & \textbf{Reddit} & \textbf{Disney} & \textbf{Books} & \textbf{Enron} & \textbf{DGraph} &  \textbf{Pokec-z}   & \textbf{Pokec-n}   & \textbf{UCSD34} & \textbf{German}   & \textbf{Credit}   & \textbf{Bail}     & \textbf{Reddit}       & \textbf{Twitter}      \\ \midrule
\textbf{\# Nodes}       & 8,405 & 10,984 & 124 & 1,418 & 13,533 & 3,700,550 & 7,659     & 6,185    & 4,132 & 1,000     & 30,000    & 18,876    & 9,892         & 47,712        \\
\textbf{\# Edges}        & \violet{407,963} & \violet{168,016} & \violet{335} & \violet{3,695} & \violet{176,987} & \violet{4,300,999} & \violet{29,476}    & \violet{21,844}  & \violet{108,383} & 24,970    & 2,174,014  & 403,977   & \textbf{\violet{1,211,748}}      & \textbf{\violet{468,697}}       \\
\textbf{\# Attributes} & 400 & 64 & 28 & 21 & 18 & 17 & 59 & 59 & 7 & 27 & 18 & 13 & 385 & 780 \\
\textbf{Avg. degree}    & 48.5 & 15.3 & 2.7 & 2.6 & 13.1 & 1.2 & 7.70 & 7.06 & 52.5 & 25.0     & 72.5     & 21.4     & 122.5        & 9.8          \\
\textbf{Real graph?}    & \cmark & \cmark & \cmark & \cmark & \cmark & \cmark & \cmark & \cmark & \cmark & \Large$\times$     & \Large$\times$     & \Large$\times$    & \cmark       & \cmark          \\\midrule
\multirow{2}{*}{\begin{tabular}[c]{@{}c@{}}\textbf{Sensitive}\vspace{-0.15cm}\\\textbf{attributes}\end{tabular}} & \multirow{2}{*}{-} & \multirow{2}{*}{-} & \multirow{2}{*}{-} & \multirow{2}{*}{-} & \multirow{2}{*}{-} & \multirow{2}{*}{-} & \multirow{2}{*}{\red{Region}} & \multirow{2}{*}{\red{Region}} & \multirow{2}{*}{\red{Gender}} & \multirow{2}{*}{\red{Gender}} & \multirow{2}{*}{\red{Age}} & \multirow{2}{*}{\red{Race}} & \multicolumn{2}{c}{\multirow{2}{*}{\textbf{\red{Political leaning}}}} \\
  & &   &  &  &  &  &  &  &  &  &  &  &      &     \\
\textbf{Attribute bias}  &- &  - & - & - & - & - & 4.3E-4 & 5.4E-4 & 5.3E-4 & 6.33E-3 & 2.46E-3 & 9.5E-4 & 2.22E-3     & 9.14E-4     \\
\textbf{Structural bias} &- & - & - & - & - & -  & 8.3E-4 & 1.03E-3 & 6.8E-4 & 1.04E-2 & 4.45E-3 & 1.1E-3 & 4.55E-4  & 6.38E-4     \\ \midrule
\multirow{2}{*}{\begin{tabular}[c]{@{}c@{}}\textbf{Anomaly}\vspace{-0.15cm} \\ \textbf{labels}\end{tabular}} & \multirow{2}{*}{\begin{tabular}[c]{@{}c@{}}\blue{Suspicious}\vspace{-0.15cm} \\ \blue{users}\end{tabular}} & \multirow{2}{*}{\begin{tabular}[c]{@{}c@{}}\blue{Banned}\vspace{-0.15cm} \\ \blue{users}\end{tabular}} & \multirow{2}{*}{\begin{tabular}[c]{@{}c@{}}\blue{Manual}\vspace{-0.15cm} \\ \blue{label}\end{tabular}} & \multirow{2}{*}{\begin{tabular}[c]{@{}c@{}}\blue{Tag of}\vspace{-0.15cm} \\ \blue{amazonfail}\end{tabular}} & \multirow{2}{*}{\begin{tabular}[c]{@{}c@{}}\blue{Spammer}\vspace{-0.15cm} \\ \blue{accounts}\end{tabular}} & \multirow{2}{*}{\begin{tabular}[c]{@{}c@{}}\blue{Overdue}\vspace{-0.15cm} \\ \blue{accounts}\end{tabular}} & \multirow{2}{*}{-}    & \multirow{2}{*}{-}    & \multirow{2}{*}{-} & \multirow{2}{*}{\begin{tabular}[c]{@{}c@{}}\blue{Credit}\vspace{-0.15cm} \\ \blue{status} \end{tabular}} & \multirow{2}{*}{\begin{tabular}[c]{@{}c@{}}\blue{Bail}\vspace{-0.15cm} \\ \blue{decision} \end{tabular}} & \multirow{2}{*}{\begin{tabular}[c]{@{}c@{}}\blue{Future}\vspace{-0.15cm} \\ \blue{default} \end{tabular}} & \multicolumn{2}{c}{\multirow{2}{*}{\textbf{\begin{tabular}[c]{@{}c@{}}\blue{Misinformation}\vspace{-0.15cm} \\ \blue{spreader}\end{tabular}}}} \\ 
& & & & & & & & & & & & & & \\
\textbf{Contamination}  & 0.103 & 0.033 & 0.048 & 0.020 & 0.004 & 0.004  & - & - & - & 0.300   & 0.221    & 0.376    & 0.137        & 0.067        \\ \midrule
\textbf{Correlation}  &- &  - & - & - & - & - &  - & - & - & 0.462   & 0.513    & 0.460    & 0.802        & 0.896        \\ 
\bottomrule
\end{tabular}
}
\vspace{-0.2cm}
\end{table*}

\textbf{Background}.
\textit{Graph Anomaly Detection} (GAD) aims to identify anomalous nodes in an input graph whose characteristics are significantly different from those of the rest nodes in the graph~\citep{GAD_DOMINANT, ding2019interactive, ding2021cross, xu2022contrastive}. 
Given that many types of real-world data, including social networks~\citep{kumar2018community, kumar2019predicting,LeeLLK21,KangLLHK21,YooLSK23}, 
recommender systems~\cite{LeeK018,KimLSK22,YangHXL22,jin2023code}, 
and cybersecurity~\citep{ lakhaAnomalyDetectionCybersecurity2022}, can be naturally represented as graphs, there has been an increasing interest in research on developing GAD methods in recent years~\citep{ma2021survey}.
By detecting anomalies in graphs, we can characterize potential threats and harmful content, enabling early warning, timely intervention, and efficient decision-making. 

With the advance of Graph Neural Networks (GNNs)~\citep{KipfW17,MNCI_ML_SIGIR,TGC_ML,NiepertAK16, zhu2021deep,liang2024survey,TGC_ML_ICLR,SharmaLNSSKK24,KimLK23,LeeL0K22,KongKJ0LPK22}, GNN-based GAD methods have increasingly attracted attention in the literature~\citep{KimLSL22,ma2021survey}.
In a nutshell, these methods usually employ GNNs to generate node embeddings that preserve both the structure and attribute information of nodes in the input graph. 
These embeddings are then utilized to reconstruct the adjacency and attribute matrices of the graph, thereby identifying anomalous nodes having high reconstruction errors.

\vspace{1mm}
\noindent\textbf{Motivation}.
\textit{Considering fairness in GAD research} is essential due to the widespread application of GAD methods in high-stakes domains, such as 
abnormal transactions~\cite{ChenQLX22}
and misinformation~\citep{chen2022combating,wang2023attacking} detection, where biased and unfair anomaly detection outcomes can have adverse effects on various aspects of our lives~\citep{dong2021individual, dong2023fairness, wu2022bias}. 
Despite the advances in GAD methods, there has been a notable lack of in-depth investigation into their ability to produce the desired results from a \textit{fairness perspective}.
The literature has demonstrated that graph mining algorithms can yield discriminatory results against \textit{sensitive attributes} (\eg, gender and political leanings) due to biases introduced
during the mining process~\citep{dong2023fairness,KangT21, wang2022training}.
Such observations raise concerns regarding the potential for existing GAD methods to produce \textit{unfair results} in detecting anomalous nodes.

However, conducting research on \textbf{Fair Graph Anomaly Detection (\ours)} is quite challenging, primarily due to the \textit{absence of comprehensive benchmark datasets} that encompass all of the graph, anomaly detection, and fairness aspects.
As depicted in Table~\ref{tab:statistics},
the existing datasets for fairness and anomaly detection research have synthetic graph structures, or lack anomaly labels or sensitive attributes.
The use of such synthetic data fails to reflect real-world properties, while the lack of anomaly labels or sensitive attributes prevents the reasonable evaluation of existing GAD methods from a fairness perspective.
Consequently, the lack of relevant datasets presents additional difficulties in developing new \ours\ methods.

\vspace{1mm}
\noindent\textbf{Our Work}.
We create two datasets that have a real-world graph structure, anomaly labels, and sensitive attributes, and then evaluate existing GAD methods in terms of accuracy and fairness by using these datasets.
Our contributions can be summarized as follows:

\begin{itemize}[leftmargin=*]
\item {\textbf{Problem Formulation:}} We define the \ours\ problem, which serves as the foundation of our investigation regarding fairness in GAD research.
\item {\textbf{Novel Datasets:}} We create two datasets from two major social media platforms, \ie, Twitter and Reddit, and analyze their crucial properties such as 
contamination
and attribute/structural biases.
\item {\textbf{Experimental Evaluation:} Under the \ours\ problem, by using our datasets, we examine the effectiveness of four state-of-the-art GAD methods}
(\dominant~\citep{GAD_DOMINANT}, \conad~\citep{xu2022contrastive}, \cola~\citep{GAD_COLA}, and VGOD~\citep{Huang0Z023}) and seven non-graph AD methods~\citep{BandyopadhyayNV20,Li22ecod,KingmaW13,BreunigKNS00,LiuTZ08,BandyopadhyayLM19} in terms of accuracy and fairness as a solution to the \ours\ problem. We also explore the impact of incorporating five fairness methods~\citep{fair_fairod,fair_hin,fair_edits,fair_fairwalk}
into the GAD methods.
\end{itemize}

To the best of our knowledge, we are the first to present comprehensive and real-world benchmark datasets that cover all of the graph, anomaly detection, and fairness aspects, which can significantly encourage follow-up studies on \ours\ research.

\section{The Proposed Problem: \textbf{\ours}}\label{sec:framework}

\textbf{Problem Definition.} The GAD problem is commonly approached as an unsupervised node classification task on a graph (\aka, network), aiming to determine whether nodes in the graph are anomalies (\aka, outliers) or not \citep{GAD_DOMINANT,GAD_COLA, xu2022contrastive}. Anomalies typically consist of a minority of the nodes in the graph. 
Let $\mathcal{G} = (\mathcal{V}, \mathcal{E}, \mathbf{X})$ represent an attributed graph, where $\mathcal{V}$ and $\mathcal{E}$ denote the sets of nodes and edges, respectively, and $\mathbf{X} \in \mathbb{R}^{n\times d}$ represents the node feature matrix, where $n$ indicates the number of nodes in the graph and $d$ indicates the number of attributes for each node.
The adjacency matrix is denoted by $\mathbf{A} \in \{0,1\}^{n\times n}$.
The anomaly labels are represented as $\mathbf{Y} \in \{0,1\}^{n}$, where a value of 1 indicates that the node is an anomaly, and the predictions of the model are denoted as $\mathbf{\hat{Y}}$. 
GAD methods aim to identify the nodes whose patterns differ significantly from the majority in terms of both attributes and a structure.
It is worth noting that since GAD is regarded as an unsupervised problem in most literature~\citep{KimLSL22,ma2021survey}.

On the other hand, the \ours\ problem extends beyond GAD by incorporating \textit{sensitive attributes} for nodes.
For instance, 
features such as age and gender, which users are usually reluctant to share, are considered sensitive attributes. 
Thus, one of the features for each node should include a sensitive attribute, which can be represented as  $\mathbf{S} \in \{0,1\}^{n}$ if the attribute is binary --
one having a sensitive attribute of 0 (\eg, male) and the other having a sensitive attribute of 1 (\eg, female). 
\ours\ methods aim to accurately detect anomalous nodes while avoiding discriminatory predictions against individuals from any specific sensitive group.

\vspace{1mm}
\noindent\textbf{Metrics.} We employ two types of metrics to analyze the performance and fairness of GAD methods. 
\textit{Performance metrics} are used to evaluate the accuracy of the GAD methods while considering the imbalanced ratio between anomaly and normal nodes.
For this purpose, the Area Under the ROC Curve (AUCROC) is widely utilized in the literature \citep{fair_edits, liu2022bond,GAD_COLA}. Additionally, we employ the Area Under the Precision-Recall Curve (AUPRC), which is more sensitive to minority labels and thus suitable for GAD. 
Higher values of these metrics indicate better model performance. 

\textit{Unfairness metrics} are used to evaluate the fairness of the GAD methods when predicting anomalies with respect to the node's sensitive attribute. 
Statistical Parity (SP)~\citep{agarwalUnifiedFrameworkFair2021,BeutelCZC17,LouizosSLWZ15} measures the difference in prediction rates for anomalies across the two node groups with different sensitive attributes, \ie,

\vspace{-0.3cm}
\begin{equation}
SP = |P(\mathbf{\hat{Y}}=1 | \mathbf{S}=0) - P(\mathbf{\hat{Y}}=1 | \mathbf{S}=1)|.
\end{equation}
\normalsize

Another fairness measure is the Equality of Odds (EOO)~\citep{agarwalUnifiedFrameworkFair2021,BeutelCZC17,LouizosSLWZ15}, which quantifies the difference in true positive rates of the method when detecting anomalies across different sensitive attributes, \ie,

\vspace{-0.3cm}
\begin{equation}
EOO = |P(\mathbf{\hat{Y}}=1 | \mathbf{S}=0, \mathbf{Y}=1) - P(\mathbf{\hat{Y}}=1 | \mathbf{S}=1, \mathbf{Y}=1)|.
\end{equation}
\normalsize
Lower values of these metrics indicate better model fairness.

\section{Data Description}
\label{sec:datasets}

\subsection{Collection Procedure}
We focus our analysis on two globally-prominent social media platforms: \textbf{Twitter} and \textbf{Reddit}. 
Both Twitter and Reddit exemplify large-scale, mainstream social media with substantial user engagement and global reach, both ranking among the top 10 most visited websites worldwide~\citep{semrushreddit,wikipediamostvisited}.
These platforms, noted for their extensive use in prior studies~\citep{jin2022towards, kang2022framework, kumar2019predicting, ma2023characterizing, shu2019defend, verma2022examining, yang2022reinforcement}, serve as rich research environments in diverse domains.

\vspace{1mm}
\noindent\textbf{Dataset Curation.}
We gathered data for the \textbf{Twitter} dataset, encompassing all historical posts, user profiles, and follower relationships of 47,712 users through the Twitter API.
This user list was sourced from~\citet{verma2022examining}, focusing on users who shared COVID-19 related tweets 
containing misinformation.
Regarding the \textbf{Reddit} dataset, we identified 110 politics-related subreddits (listed in \textbf{\autoref{sec:pol_subreddits}}). We used the Pushshift API
to retrieve all historical posts from these subreddits and identified users engaged in the discussions within them. From these participants, we randomly sampled users and collected all their historical posts since their account creation. 
The collection of publicly available datasets was deemed exempt from review by the Institutional Review Board.

In both datasets, we define the \textbf{political leaning of users} as the \textbf{sensitive attribute}. The  \textbf{anomaly label} represents whether a user spreads \textbf{real-news or misinformation}~\citep{sakketouFACTOIDNewDataset2022}.
The correlation between political leanings and the spread of misinformation has been extensively documented in prior studies~\citep{cohen2020correct,lawson2022pandemics,GuptaDPMD23}.
To assign these labels,
we utilize
the FACTOID dataset~\citep{sakketouFACTOIDNewDataset2022}, which furnishes lists of online news outlet domains: 1,577 for misinformation and 571 for real news, and 142 left-leaning and 777 right-leaning domains.

Employing a methodology similar to~\citep{sakketouFACTOIDNewDataset2022}, 
we classify hyperlinks based on their domains, categorizing them as left/right-leaning and real news/misinformation. 
Consequently, users receive a sensitive attribute value of 1 if they post more links from right-leaning sites than left-leaning ones, and 0 vice versa. 
Similarly, users receive an anomaly label value of 1 if they share more misinformation links than real news links.
However, it's important to note that the criterion mentioned earlier for assigning the sensitive attribute and anomaly label is merely one possible approach. Our \ours\ datasets provide the flexibility to set different thresholds. For example, users might receive a sensitive attribute value of 1 if they engage with the content from a specific political ideology more frequently (\eg, $\ge5$) than others, and 0 otherwise.

Furthermore, we established the \textbf{graph structure} in both datasets.
For Reddit, we constructed the graph by connecting two users who posted to the same subreddit within a 24-hour window. This method creates an undirected edge between users, reflecting the non-hierarchical nature of interactions within the subreddit.
This approach draws from prior research indicating that users interacting within the same online community in close temporal proximity are likely to be aware of each other's posts or share similar topical interests~\citep{krohn2022subreddit, waller2019generalists}. We employed Sentence Transformers ~\citep{reimersSentenceBERTSentenceEmbeddings2019} to generate embeddings from users' post histories. We then computed the average of a user's post embeddings and combined it with their sensitive attribute to derive the node feature in our graph.

For Twitter, 
we established a directed edge from user $A$ to user $B$ if $A$ follows $B$. 
Utilizing the M3 System~\citep{wang2019demographic}, a comprehensive demographic inference framework trained on extensive Twitter data, we inferred user demographic information, including age group ($\leq$18, 19-29, 30-39, $\geq$40), gender, and organization account status, based on user profiles and historical tweets.
We also collected data on favorites and account verification status.
Users' post histories were retrieved and embedded using a multilingual model~\citep{reimersMakingMonolingualSentence2020}. We then computed the average of a user's post embeddings and concatenated them with the above user information to form the node features. 
Finally, for both datasets, we retained the largest connected component of nodes as the final graph structures. 

\subsection{Dataset Statistics}

\autoref{tab:statistics} provides an overview of the basic statistics and
the following key properties:
(1) \textbf{correlation} indicates the correlation coefficient between sensitive attributes and anomaly labels;
(2) \textbf{attribute bias}~\citep{fair_edits} employs the Wasserstein-1 distance~\citep{villani2021topics} to compare the distribution of node attributes between anomalies and non-anomalies;
(3) \textbf{structural bias}~\citep{fair_edits} uses the Wasserstein-1 distance~\citep{villani2021topics} while comparing adjacency matrices based on a two-hop neighborhood between them; and 
(4) \textbf{contamination} represents the proportion of anomaly nodes in the dataset. 

For attribute bias, let $\mathbf{X}_{norm} \in \mathbb{R}^{N\times M}$ represent the normalized attribute matrix of an input graph, where $N$ and $M$ denote the numbers of nodes and attributes, respectively.
The attribute bias $b_{attr}$, given $\mathbf{X}_{norm}$, is calculated as follows~\citep{fair_edits}:

\vspace{-0.3cm}
\small
\begin{equation}
b_{attr} = \frac{1}{M}\sum_{m=1}^M W(pdf(\mathcal{X}^0_m), pdf(\mathcal{X}^1_m)),    
\end{equation}
\normalsize
where $\mathcal{X}^0_m$ (resp. $\mathcal{X}^1_m$) denote the $m$-th attribute value sets for nodes with sensitive attributes of 0 (resp. 1).
We partition the attributes of all nodes as $\{(\mathcal{X}^0_1,\mathcal{X}^1_1),(\mathcal{X}^0_2,\mathcal{X}^1_2),\cdots,(\mathcal{X}^0_M,\mathcal{X}^1_M)\}$. 
Also, $W$ and $pdf$ denote the Wasserstein-1 distance~\citep{villani2021topics} between two distributions and the probability density function for a set of values, respectively.

For structural Bias, we denote a normalized adjacency matrix with re-weighted self-loops as $\mathbf{P}_{norm} = \alpha \mathbf{A}_{norm} + (1-\alpha)\mathbf{I}$, where $\mathbf{A}_{norm}$ and $\mathbf{I}$ represent the symmetric normalized adjacency matrix and the identity matrix, respectively; $\alpha$ is a hyperparameter ranging from 0 to 1.
The propagation matrix is defined as $\mathbf{M}_H = \sum_{h=1}^H \beta^h \mathbf{P}_{norm}^h$, where $H$ and $\beta$ indicate the number of hops for the propagation measurement and the discount factor reducing the weight of propagation from neighbors with higher hops, respectively.
Given $\mathbf{M}_H$, the structural bias $b_{struc}$ is calculated~\citep{fair_edits}:

\vspace{-0.3cm}
\small
\begin{equation}
b_{struc} = \frac{1}{M}\sum_{m=1}^M W(pdf(\mathcal{R}^0_m), pdf(\mathcal{R}^1_m)),
\end{equation}
\normalsize
where $\mathbf{R}=\mathbf{M}_H \mathbf{X}_{norm}$ represents the reachability matrix. 
Here, $\mathcal{R}^0_m$ and $\mathcal{R}^1_m$ represent the $m$-th attribute value sets in $\mathbf{R}$ for nodes with sensitive attributes of 0 and 1, respectively.

\vspace{1mm}
\noindent\textbf{Key Characteristics.} We summarize the key differences between \ours\ and the synthetic datasets, \ie, German, Credit, and Bail.

First, our datasets show a strong link (\ie, \textbf{correlation}) between sensitive attributes and anomalies. 
This supports previous studies~\citep{grinberg2019fake,GuptaDPMD23, lawson2022pandemics} on the correlation between political leanings and the spread of misinformation. As a result, a naive approach to infer anomalies based on the sensitive attributes of nodes could result in high accuracy in our datasets. 
However, this implies that the approach harms fairness by preserving the inherent correlations. 
Furthermore,
since such correlations in a dataset can leak into the graph structure and non-sensitive attributes~\citep{fair_edits,WangZDCLD22},
GAD methods 
have the potential to amplify the aforementioned biases.

In addition, our datasets present varying \textbf{graph structures} that are shaped by the features of social media platforms. 
On Reddit, users engage in numerous subreddits, resulting in a denser graph compared to the synthetic ones.
In contrast, Twitter's graph is sparser than the synthetic ones due to its directed edges that represent user-following relationships, leading to a lower average degree. 

The synthetic datasets were initially created for non-graph AD, where synthetic edges were formed by linking nodes using the Minkowski distance, without considering actual user behavior.
As a consequence, the inductive biases of GAD methods may not be as applicable, since they often rely on assumptions that anomalies differ from their neighboring nodes~\citep{GAD_COLA, xu2022contrastive}.
In contrast, our datasets exhibit less \textbf{structural bias} than the synthetic ones due to its origins in actual user behavior.
This difference is because users with distinct properties may still be connected in social networks. 
This is supported by the average similarity between users connected by edges of 43\% and 44\% for our Reddit and Twitter, respectively, which contrasts with the thresholds used to create the synthetic edges for German, Credit, and Bail at 80\%, 70\%, and 60\%, respectively~\citep{fair_edits}. 

Lastly, our Twitter dataset exhibits the lowest \textbf{attribute bias} out of our \ours\ and the synthetic datasets. Additionally, it includes a larger number of attributes than other datasets. 
According to~\citet{zimekSurveyUnsupervisedOutlier2012}, such properties (\ie, low attribute bias and high dimensionality of attributes) are known to make anomaly detection more challenging, which will be demonstrated in Section~\ref{sec:results}.

\subsection{Why Do Our Datasets Matter and Suit the \ours\ Problem?}
We recognize a multitude of data collection methods and sources that can be considered for tackling the \ours\ problem. 
Amongst design choices, the rationale behind our selections is as follows.

Addressing fairness concerns in politically biased misinformation detection poses a series of practical implications, as political bias can reinforce confirmation biases, treat news sources unequally, and impede just categorization. 
Misclassifying minority groups as misinformation spreaders can amplify biases and stereotypes, potentially deepening existing divisions through algorithmic misuse. 
Therefore, prioritizing fairness builds trust in the detection process and fosters a more equitable information environment.

Moreover, our chosen topics stem from extensive research on misinformation propagation in \textit{COVID-19 and politics}~\citep{bin2021covid,HeZSRYK21,LuAKZ23,MicallefHKAM20}. 
In politics, this is crucial due to the potential for polarization and ideological divisions stemming from such misinformation, influencing public discourse and decision-making processes. 
Regarding COVID-19, unverified claims or inaccurate information about the virus, prevention methods, and treatments can prompt misguided actions that worsen the pandemic's impact and impede effective response efforts.
We would like to clarify the scope of our datasets, specifically regarding misinformation about COVID-19 (for the Twitter dataset) and politics-related misinformation (for the Reddit dataset). 
As such, our datasets do not fully represent the broader populations on Twitter and Reddit.

Lastly, we carefully considered the suitability of our collection process by conducting human verification of the labeling methodology.  
We randomly sampled 1,000 posts in the Reedit dataset containing URLs with either left or right political leaning. 
Three annotators, with no conflict of interest, assessed users' political leanings based on their sharing behavior. 
They also reviewed the related content and posts, matching the sentiment expressed by users with the political leanings of the posts. Annotators achieved 99.8\% agreement with a Fleiss’ Kappa~\cite{fleiss2013statistical} of 0.793, indicating substantial agreement. Using majority voting, we found that in 99.6\% of cases, users supported the political leaning associated with the post's URL. This empirical evidence suggests that shared website hyperlinks substantially help classify users' political leanings.
The ethical considerations for data collection, including user privacy and bias perpetuation, are discussed in Section~\ref{sec:ethics}.

\vspace{-0.2cm}
\section{Evaluation}\label{sec:evaulation}
In this section, we conduct comprehensive experiments to achieve the following evaluation goals:
\begin{enumerate}[leftmargin=*]
    \item We assess the efficacy of current GAD methods on our \ours\ datasets 
to examine their performance 
in terms of both fairness and accuracy (\textbf{Section~\ref{sec:results}}).
    \item We evaluate 
the impact of integrating fairness methods into GAD methods, consequently revealing the trade-off space between accuracy and fairness (\textbf{Section~\ref{sec:results}}).
    \item We delve deeper into the \ours\ problem to gain insights into the effects of sensitive attributes (\textbf{Section~\ref{sec:alter_sensitive}}).
\end{enumerate}

\subsection{Experimental Settings}\label{sec:setting}

\begin{table*}[t]
\centering
\caption{Performance and fairness results of GAD methods on our original and debiased datasets. 
$\uparrow$ means higher values are better; $\downarrow$ means lower values are better; `o.o.m' denotes out of memory.}
\label{tab:gad_debiaser}
\vspace{-0.25cm}
\resizebox{\textwidth}{!} {
 \renewcommand{\arraystretch}{1.1}
 \setlength{\tabcolsep}{2pt} 
\begin{tabular}{c|ccc|ccc|ccc|ccc}
\multicolumn{13}{c}{\large\textbf{(a) Twitter Dataset}} \\ 
\toprule
\multicolumn{1}{c|}{\textbf{Methods}} & \multicolumn{3}{c|}{\textbf{CoLA}} & \multicolumn{3}{c|}{\textbf{CONAD}} & \multicolumn{3}{c|}{\textbf{DOMINANT}} & \multicolumn{3}{c}{\textbf{VGOD}} \\ \cmidrule(lr){3-3}\cmidrule(lr){6-6}\cmidrule(lr){9-9}\cmidrule(lr){12-12}
\multicolumn{1}{c|}{\textbf{Debiasers}} & \textbf{\boldsymbol{$\times$}} & \textbf{EDITS} & \textbf{FairWalk} & \textbf{\boldsymbol{$\times$}} & \textbf{EDITS} & \textbf{FairWalk} & \textbf{\boldsymbol{$\times$}} & \textbf{EDITS} & \textbf{FairWalk} & \textbf{\boldsymbol{$\times$}} & \textbf{EDITS} & \textbf{FairWalk} \\ \midrule
\textbf{AUCROC} ($\uparrow$) &	 0.443±0.006	& 0.452±0.013 & \textbf{0.488±0.006} 	& 0.558±0.007 	 & \textbf{0.704±0.001}  & 0.536±0.009 & 0.560±0.007 & \textbf{0.704±0.001} &  0.535±0.009 & 0.736±0.006 & \textbf{0.823±0.032} &  0.602±0.003 \\ 
\textbf{AUPRC} ($\uparrow$) &	 0.052±0.001	& 0.053±0.002	& \textbf{0.062±0.002}	& 0.087±0.001 	 & \textbf{0.173±0.001} & 0.085±0.005  &  0.088±0.001 & \textbf{0.173±0.001} & 0.085±0.005 &  0.159±0.009 & \textbf{0.241±0.020} & 0.091±0.001  \\ \midrule
\textbf{SP} ($\downarrow$)  &	 0.028±0.003	& \textbf{0.007±0.006}	& 0.008±0.006	& 0.038±0.006 	 & 0.289±0.004  & \textbf{0.011±0.004} & 0.040±0.006 & 0.289±0.003 & \textbf{0.012±0.004}  & 0.124±0.012 & 0.172±0.073 & \textbf{0.098±0.003}   \\ 
\textbf{EOO} ($\downarrow$)  &	 0.023±0.012	& 0.009±0.005	& \textbf{0.001±0.001} 	& 0.044±0.003 	 & 0.278±0.004 & \textbf{0.013±0.002} & 0.044±0.003 & 0.278±0.003 & \textbf{0.013±0.002} & 0.111±0.021 & 0.144±0.085 &  \textbf{0.052±0.004} \\ 
\bottomrule
\end{tabular}}
\resizebox{\textwidth}{!}{
 \renewcommand{\arraystretch}{1.1}
 \setlength{\tabcolsep}{3pt} 
\begin{tabular}{c|ccc|ccc|ccc|ccc}
\multicolumn{13}{c}{\vspace{-0.2cm}} \\ 
\multicolumn{13}{c}{\large\textbf{(b) Reddit Dataset}} \\ 
\toprule
\multicolumn{1}{c|}{\textbf{Methods}} & \multicolumn{3}{c|}{\textbf{CoLA}} & \multicolumn{3}{c|}{\textbf{CONAD}} & \multicolumn{3}{c|}{\textbf{DOMINANT}} & \multicolumn{3}{c}{\textbf{VGOD}} \\ 
\cmidrule(lr){3-3}\cmidrule(lr){6-6}\cmidrule(lr){9-9}\cmidrule(lr){12-12}
\multicolumn{1}{c|}{\textbf{Debiasers}} & \textbf{\boldsymbol{$\times$}} & \hspace{0.5cm}\textbf{EDITS}\hspace{0.5cm} & \textbf{FairWalk} & \textbf{\boldsymbol{$\times$}} & \hspace{0.5cm} \textbf{EDITS} \hspace{0.5cm} & \textbf{FairWalk} & \textbf{\boldsymbol{$\times$}} & \hspace{0.5cm}\textbf{EDITS}\hspace{0.5cm} & \textbf{FairWalk} & \textbf{\boldsymbol{$\times$}} & \hspace{0.5cm}\textbf{EDITS}\hspace{0.5cm} & \textbf{FairWalk} \\ \midrule
\textbf{AUCROC} ($\uparrow$) & 0.453±0.014	& o.o.m	& \textbf{0.502±0.004}	& \textbf{0.608±0.001} & o.o.m & 0.517±0.024 & \textbf{0.608±0.001} & o.o.m & 0.518±0.023 & \textbf{0.721±0.009} & o.o.m & 0.673±0.002 \\ 
\textbf{AUPRC} ($\uparrow$) &	 0.032±0.018 	&	o.o.m & \textbf{0.140±0.005}	& \textbf{0.200±0.001} & o.o.m & 0.149±0.015 &  \textbf{0.200±0.001} & o.o.m & 0.150±0.016 &  \textbf{0.394±0.024} & o.o.m & 0.284±0.001 \\ \midrule
\textbf{SP} ($\downarrow$) &	 0.035±0.027	& o.o.m	& \textbf{0.006±0.004}	& 0.132±0.001 & o.o.m & \textbf{0.025±0.017} & 0.133±0.002 & o.o.m & \textbf{0.021±0.015} & 0.427±0.058 & o.o.m & \textbf{0.317±0.005} \\ 
\textbf{EOO} ($\downarrow$) &	 0.177±0.014	& o.o.m	& \textbf{0.003±0.003} & 0.055±0.002 & o.o.m & \textbf{0.028±0.018} & 0.057±0.003  & o.o.m & \textbf{0.025±0.017} & 0.472±0.063 & o.o.m &  \textbf{0.295±0.006}\\ 
\bottomrule
\end{tabular}}
\vspace{-0.25cm}
\end{table*}

\vspace{1mm}
\noindent\textbf{GAD Methods.} 
We employ \textit{four GAD methods}, \ie, \dominant~\citep{GAD_DOMINANT}, \conad~\citep{xu2022contrastive},  \cola~\citep{GAD_COLA}, and VGOD~\citep{Huang0Z023}. 
Our goal is to present new datasets and to investigate their properties and applicability in terms of graphs, fairness, and anomaly detection aspects. 
Therefore, we have chosen representative or state-of-the-art GAD methods rather than using all GAD methods.
Beyond graph-based methods, we also examine 
\textit{five non-graph AD methods} (\ie, DONE~\citep{BandyopadhyayNV20}, AdONE~\citep{BandyopadhyayNV20}, ECOD~\citep{Li22ecod}, VAE~\citep{KingmaW13}, and ONE~\citep{BandyopadhyayLM19}), and \textit{two heuristic methods} (\ie, LOF~\citep{BreunigKNS00} and IF~\citep{LiuTZ08}) in \textbf{Appendix~\ref{app:additional_baselines}}.

\textit{\dominant}~\citep{GAD_DOMINANT} uses GCNs to obtain node embeddings, which are then used in other GCNs to reconstruct the attribute and the adjacency matrices. 
By measuring the errors between the original and decoded matrices, anomalies are detected. 
Under the premise that anomalous nodes are more difficult to encode than normal nodes, it ranks nodes based on their reconstruction errors. 
The top nodes with high reconstruction errors are identified as anomalies.

\textit{\conad} \citep{xu2022contrastive} 
incorporates human knowledge about different anomaly types into detecting anomalies through knowledge modeling. Synthetic anomalies are introduced into the graph for self-supervised learning via a contrastive loss.
Similar to \dominant, 
the reconstruction error is then used to label nodes as anomalies. 

\textit{\cola} \citep{GAD_COLA} employs self-supervised learning with pairs of a contrastive node and local neighborhood obtained by random walks. 
This subsampling strategy assumes that anomalies and their neighborhoods differ from normal nodes and their neighborhoods. 
The learned model compares all nodes in the graph with their neighborhoods via positive and negative pairs to identify nodes in which the model cannot distinguish between positive and negative pairs, 
which are then predicted as anomalies.

\textit{\vgod}~\citep{Huang0Z023} is a recent GAD method that focuses on structural outliers. By using a novel variance-based method, VGOD samples positive and negative edges in the graph to capture the information of node neighborhoods in its anomaly detection model through a contrastive loss between them.

\vspace{1mm}
\noindent\textbf{Fairness Methods.}
We employ \textit{five fairness methods} that are applicable to the GAD problem: (1) fairness regularizers:  \fairod~ \citep{fair_fairod}, \correlation~\citep{fair_fairod}, and \hin~\citep{fair_hin}; (2) graph debiasers: \edits~\citep{fair_edits} and \fairwalk~\citep{fair_fairwalk}. 
These methods are used to enhance the fairness of GAD methods by reducing fairness metrics while minimizing the impact on model performance. 
Detailed equations for fairness regularizers can be found in \textbf{Appendix~\ref{sec:fairness_equation}}.

\textit{\fairod} \citep{fair_fairod} originally proposed for unsupervised, non-graph AD, focuses on improving fairness through EOO.
It states that solely improving SP can lead to models lazily predicting the same number of anomalies for each sensitive attribute. 
Therefore, \fairod\ introduces two losses $\mathcal{L}_{FairOD}$ for SP, which reduces the sum of reconstruction errors, and $\mathcal{L}_{ADCG}$ for EOO, which penalizes the fair model for ranking nodes differently from the original model. 

The \textit{\correlation} regularizer $\mathcal{L}_{Corr}$, derived from the \fairod\ implementation,
measures the correlation between sensitive attributes and node representation errors by using the cosine rule.
This ensures that nodes are encoded to achieve similar accuracy, regardless of any sensitive attributes. 

\textit{\hin} \citep{fair_hin} is another regularizer for fairness representation learning. While it originally intends for heterogeneous information networks, the loss function can be adapted to GAD to reduce the same SP fairness metric. 
The loss $\mathcal{L}_{HIN}$ penalizes the difference in prediction rates between sensitive attribute groups for both anomalies and non-anomalies. \citet{fair_hin} introduces another function that reduces EOO, but requires labels.
Thus, we use $\mathcal{L}_{ADCG}$ from the \fairod\ regularizer as a replacement.

There are other fairness methods designed for debiasing graphs. 
\textit{\edits} \citep{fair_edits} takes the graph and node features as input and employs gradient descent to learn a function that debiases them by reducing the estimated Wasserstein distance between attribute dimensions and the node label. This results in modifications to the adjacency matrix (by removing or adding edges) as well as the node feature matrix, while keeping the node labels unchanged. 
In~\citep{fair_edits}, the authors claim that these modifications result in a graph with reduced bias while maintaining performance for downstream tasks. 

\textit{\fairwalk} \citep{fair_fairwalk} 
aims to generate fairer node embeddings of a graph without relying on node features, only using sensitive attributes. 
Based on node2vec, 
It modifies random walks in the graph by considering the sensitive attribute of the nodes at each step of the random walk. 
This ensures that the nodes with a minority sensitive attribute are explored more, leading to fairer representations. 
We use these embeddings as node features for GAD methods.

\vspace{1mm}
\noindent\textbf{Implementation Details}. 
We used the PyGOD\footnote{\url{https://github.com/pygod-team/pygod}} \citep{liu2022pygod} implementation of \dominant, \conad, \cola, and \vgod\ methods.
To incorporate the fairness regularizer methods (\ie, \fairod, \hin, and \correlation) into GAD methods, we made appropriate modifications to the code sections in PyGOD:
(1) $\mathcal{L} = \mathcal{L}_o + \lambda\mathcal{L}_{FairOD} + \gamma\mathcal{L}_{ADCG}$ for \fairod; (2) $\mathcal{L} = \mathcal{L}_o + \lambda\mathcal{L}_{HIN} + \gamma\mathcal{L}_{ADCG}$ for \hin; and (3) $\mathcal{L} = \mathcal{L}_o + \lambda\mathcal{L}_{corr}$ for \correlation, where $\mathcal{L}_o$ denotes the original loss of the GAD method, and $\lambda$ and $ \gamma$ are hyperparameters. 
All the experiments were conducted with the NVIDIA DGX-1 system with 8 NVIDIA TESLA V100 GPUs. 
\ul{Each experiment was repeated twenty times to ensure the robustness and reliability of the results}. 
For the full reproducibility of our research, we provide complete implementation details in \textbf{Appendix~\ref{sec:implementation}}.

\subsection{Accuracy vs. Fairness}\label{sec:results}

We conducted extensive experiments to compare the different variants of GAD methods with/without fairness methods: (1) GAD methods on the original \ours\ datasets; (2) GAD methods on the debiased \ours\ datasets generated through \fairwalk\ or \edits; (3) GAD methods with fairness regularizers, \ie, \fairod, \correlation, and \hin, on the original \ours\ datasets.

\vspace{1mm}
\noindent\textbf{Using GAD Methods (without Fairness Methods).}
We evaluate the performance and fairness of GAD methods without incorporating any fairness methods, as shown in the `without fairness methods' 
columns (\ie, $\times$) in Table~\ref{tab:gad_debiaser}.
In general, the accuracy (\ie, AUCROC and AUPRC) of GAD methods on Reddit tends to be higher than their accuracy on Twitter.
However, we found the suboptimal performance of existing GAD methods in terms of accuracy, which may be influenced by several factors. 
One possible reason is that our datasets manifest less structural bias than existing synthetic datasets, which may result in the limited performance of GAD methods due to their prevalent reliance on graph homophily. 

Furthermore, we observe that striving for higher accuracy via existing GAD methods adversely affects their fairness, which leads to higher SP and EOO. 
That is, they show worse SP and EOO on Reddit than on Twitter.
Considering that the attribute bias of Reddit is significantly larger than that of Twitter while their structural biases are similar (see Table~\ref{tab:statistics}), we attribute the results of high SP and EOO on Reddit to its substantial attribute bias.

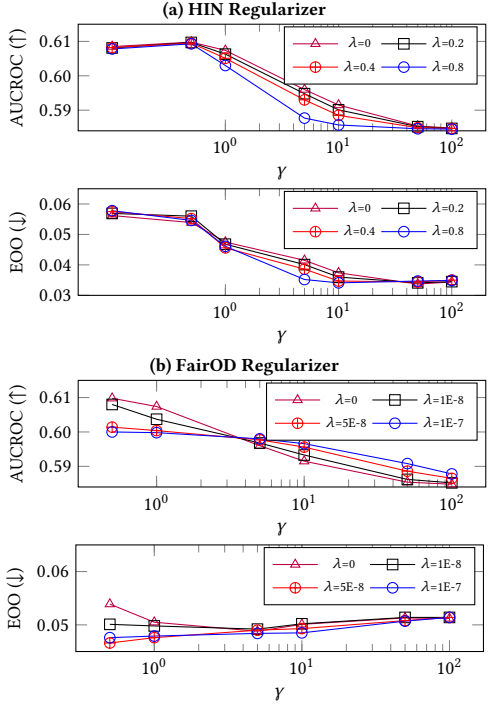
\begin{figure}[t]
\footnotesize
\centering \textbf{(a) HIN Regularizer}\\
\begin{tikzpicture}
\begin{axis}[
height=3cm, width=7cm,
xtick={0, 1, 10, 100},
legend columns=2,
xmode=log,
ylabel=AUCROC ($\uparrow$),xlabel=$\gamma$,
ymin=0.584, ymax=0.615,
y tick label style={/pgf/number format/.cd,fixed,fixed zerofill,precision=2,/tikz/.cd},]
\addplot[color=purple,mark=triangle,]
coordinates {(0, 0.6085)(0.1,0.6085)(0.5,0.6098)(1,0.6074)(5,0.596)(10,0.5915)(50,0.5854)(100,0.5848)};\addlegendentry{\tiny$\lambda$=0}%
\addplot[color=black,mark=square,]
coordinates {(0, 0.6079)(0.1,0.6081)(0.5,0.6097)(1,0.6063)(5,0.5948)(10,0.5901)(50,0.5852)(100,0.5847)};\addlegendentry{\tiny$\lambda$=0.2} 
\addplot[color=red,mark=oplus,]
coordinates {(0, 0.608)(0.1,0.6082)(0.5,0.6095)(1,0.6051)(5,0.593)(10,0.5885)(50,0.5849)(100,0.5847)};\addlegendentry{\tiny$\lambda$=0.4}%
\addplot[color=blue,mark=o,]
coordinates {(0, 0.607)(0.1,0.6078)(0.5,0.6093)(1,0.603)(5,0.5877)(10,0.5857)(50,0.5846)(100,0.5845)};\addlegendentry{\tiny$\lambda$=0.8} %
\end{axis}
\end{tikzpicture}
\begin{tikzpicture}
\begin{axis}[
height=3cm, width=7cm,
xtick={0, 1, 10, 100},
legend columns=2,
xmode=log,
ylabel=EOO ($\downarrow$),xlabel=$\gamma$,
ymin=0.03, ymax=0.065,
scaled y ticks = false,
y tick label style={/pgf/number format/.cd,fixed,fixed zerofill,precision=2,/tikz/.cd},]
\addplot[color=purple,mark=triangle,]
coordinates {(0, 0.0557)(0.1,0.0562)(0.5,0.0539)(1,0.0475)(5,0.0415)(10,0.0374)(50,0.0337)(100,0.0344)};\addlegendentry{\tiny$\lambda$=0}
\addplot[color=black,mark=square,]
coordinates {(0, 0.0565)(0.1,0.0569)(0.5,0.056)(1,0.0468)(5,0.0401)(10,0.036)(50,0.0341)(100,0.0345)};\addlegendentry{\tiny$\lambda$=0.2}
\addplot[color=red,mark=oplus,]
coordinates {(0, 0.0563)(0.1,0.0575)(0.5,0.0552)(1,0.0456)(5,0.0386)(10,0.0347)(50,0.0344)(100,0.035)};\addlegendentry{\tiny$\lambda$=0.4}
\addplot[color=blue,mark=o,]
coordinates {(0, 0.0581)(0.1,0.0578)(0.5,0.0546)(1,0.0461)(5,0.0352)(10,0.0341)(50,0.0347)(100,0.0349)};\addlegendentry{\tiny$\lambda$=0.8} 
\end{axis}
\end{tikzpicture}
\\
\textbf{(b) FairOD Regularizer}\\
\begin{tikzpicture}
\begin{axis}[
height=3cm, width=7cm,
xtick={0, 1, 10, 100},
legend columns=2,
xmode=log,
ylabel=AUCROC ($\uparrow$),xlabel=$\gamma$,
ymin=0.584, ymax=0.615,
y tick label style={/pgf/number format/.cd,fixed,fixed zerofill,precision=2,/tikz/.cd},]
\addplot[color=purple,mark=triangle,]
coordinates {(0, 0.6085)(0.5,0.6098)(1,0.6074)(5,0.596)(10,0.5915)(50,0.5854)(100,0.5848)};\addlegendentry{\tiny$\lambda$=0}%
\addplot[color=black,mark=square,]
coordinates {(0, 0.6092)(0.5,0.6080)(1,0.6037)(5,0.5968)(10,0.5933)(50,0.5862)(100,0.5852)};\addlegendentry{\tiny$\lambda$=1E-8} %
\addplot[color=red,mark=oplus,]
coordinates {(0, 0.6027)(0.5,0.6014)(1,0.6004)(5,0.5977)(10,0.5956)(50,0.5886)(100,0.5865)};\addlegendentry{\tiny$\lambda$=5E-8} %
\addplot[color=blue,mark=o,]
coordinates {(0, 0.6005)(0.5,0.6)(1,0.5998)(5,0.598)(10,0.5966)(50,0.5908)(100,0.5878)};\addlegendentry{\tiny$\lambda$=1E-7}  
\end{axis}
\end{tikzpicture}
\begin{tikzpicture}
\begin{axis}[
height=3cm, width=7cm,
xtick={0, 1, 10, 100},
legend columns=2,
xmode=log,
ylabel=EOO ($\downarrow$),xlabel=$\gamma$,
ymin=0.045, ymax=0.065,
ytick={0.05, 0.06},
scaled y ticks = false,
y tick label style={/pgf/number format/.cd,fixed,fixed zerofill,precision=2,/tikz/.cd},]
\addplot[color=purple,mark=triangle,]
coordinates {(0, 0.0557)(0.5,0.0539)(1,0.0505)(5,0.0488)(10,0.0501)(50,0.0513)(100,0.0512)};\addlegendentry{\tiny$\lambda$=0}
\addplot[color=black,mark=square,]
coordinates {(0, 0.0515)(0.5,0.0501)(1,0.0498)(5,0.0492)(10,0.0502)(50,0.0514)(100,0.0514)};\addlegendentry{\tiny$\lambda$=1E-8}
\addplot[color=red,mark=oplus,]
coordinates {(0, 0.0480)(0.5,0.0466)(1,0.0476)(5,0.049)(10,0.0493)(50,0.0508)(100,0.0514)};\addlegendentry{\tiny$\lambda$=5E-8}
\addplot[color=blue,mark=o,]
coordinates {(0, 0.0471)(0.5,0.0476)(1,0.0479)(5,0.0484)(10,0.0485)(50,0.0507)(100,0.0514)};\addlegendentry{\tiny$\lambda$=1E-7} 
\end{axis}
\end{tikzpicture}
\vspace{-0.2cm}
\caption{Changes in AUCROC and EOO for different values of $\lambda$ (HIN or FairOD factor) and $\gamma$ (ADCG factor) for \conad\ method with \hin\ and \fairod\ regularizers on Reddit.}\label{fig:fairod_hin}
\vspace{-0.25cm}
\end{figure}
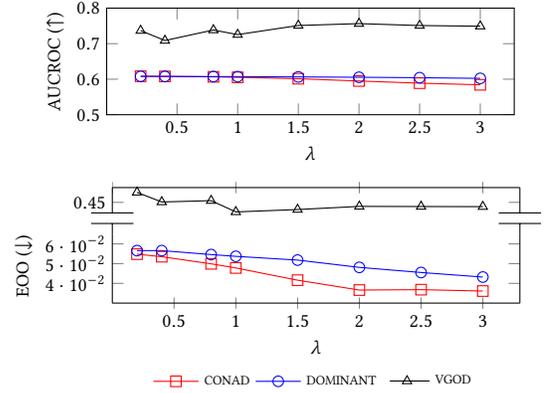
\begin{figure}[t]
\footnotesize
\begin{tikzpicture}
\begin{axis}[
height=3cm,width=7cm, 
xtick={0.5, 1.0, 1.5, 2.0, 2.5, 3.0},
ylabel=AUCROC ($\uparrow$),xlabel=$\lambda$,
ymin=0.5, ymax=0.8,
]
\addplot[color=red,mark=square,]
coordinates {(0.2, 0.6083)(0.4,0.6078)(0.8,0.6066)(1,0.6052)(1.5,0.6013)(2,0.5949)(2.5,0.5888)(3,0.5842)};
\addplot[color=blue,mark=o,]
coordinates {(0.2, 0.608)(0.4,0.6079)(0.8,0.6074)(1,0.6072)(1.5,0.6064)(2,0.6054)(2.5,0.6042)(3,0.6023)};
\addplot[color=black,mark=triangle,]
coordinates {(0.2, 0.7368)(0.4,0.7088)(0.8,0.7384)(1,0.7255)(1.5,0.7511)(2,0.7564)(2.5,0.751)(3,0.7491)};
\end{axis}
\end{tikzpicture}
\begin{tikzpicture}
\begin{groupplot}[
    group style={
        group name=my fancy plots,
        group size=1 by 2,
        xticklabels at=edge bottom,
        vertical sep=0pt,        
    },
    width=7cm,
    xmin=0, xmax=,        
]
\nextgroupplot[
xmin=0, xmax=3.3,
ymin=0.3, ymax=0.52,
ytick={0.45},
axis x line=top, 
axis line style={-},
axis y discontinuity=parallel,
height=2.2cm, 
scaled y ticks = false,
legend style={at={(0.78,-2.8)},anchor=north,draw=none,},
]
\addplot[color=black,mark=triangle,]
coordinates {(0.2, 0.4974)(0.4,0.4518)(0.8,0.4581)(1,0.4048)(1.5,0.416)(2,0.4309)(2.5,0.4303)(3,0.4297)};\addlegendentry{\tiny VGOD}%
\nextgroupplot[
height=2.5cm,
xlabel=$\lambda$,
ylabel=\vspace{-0.2cm} EOO ($\downarrow$),
ymin=0.03, ymax=0.065,
ytick={0.04, 0.05,0.06},
xmin=0, xmax=3.3,
xtick={0.5, 1.0, 1.5, 2.0, 2.5, 3.0},
scaled y ticks = false,
axis x line=bottom,
axis line style={-},
legend columns=2,
legend style={at={(0.38,-0.89)},anchor=north,draw=none,},
]
\addplot[color=red,mark=square,]
coordinates {(0.2, 0.0548)(0.4,0.0535)(0.8,0.0499)(1,0.0478)(1.5,0.0416)(2,0.0366)(2.5,0.0368)(3,0.0361)};\addlegendentry{\tiny CONAD};%
\addplot[color=blue,mark=o,]
coordinates {(0.2, 0.0566)(0.4,0.0566)(0.8,0.0546)(1,0.0537)(1.5,0.0518)(2,0.0481)(2.5,0.0455)(3,0.0432)};\addlegendentry{\tiny DOMINANT};%
\end{groupplot}
\end{tikzpicture}
\vspace{-0.25cm}
\caption{Changes in AUCROC and EOO for different values of $\lambda$ (Correlation factor) for \conad, \dominant, and \vgod\ methods with \correlation\ regularizer on Reddit.}\label{fig:correlation}
\vspace{-0.4cm}
\end{figure}
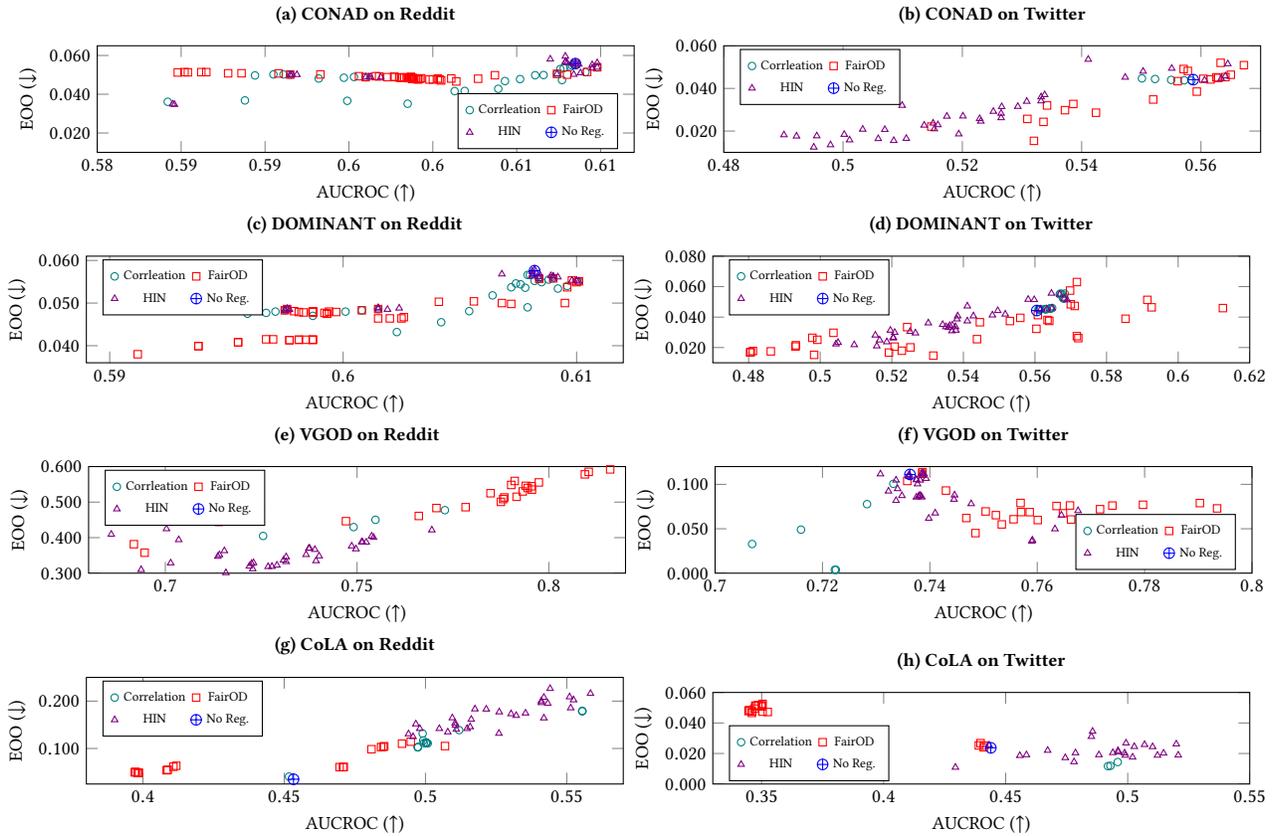
\begin{figure*}[t]
\footnotesize
\centering
\begin{tikzpicture}
    \centering
    \begin{axis}[
        xlabel={AUCROC ($\uparrow$)},
        ylabel={EOO ($\downarrow$)},
        title={\textbf{(a) CONAD on Reddit}},
        height=3cm, width=0.49\linewidth,
        ymin=0.01, ymax=0.065,
        xmin=0.58, xmax=0.612,
        scaled y ticks = false,
        legend columns=2,
        legend pos=south east,
        y tick label style={/pgf/number format/.cd,fixed,fixed zerofill,precision=3,/tikz/.cd},
        ]
        \addplot[only marks, color=teal,mark=o,mark options={scale=0.7}] file {tradeoff-conad_correlation_reddit.txt};\addlegendentry{\tiny Corrleation}        
        \addplot[only marks,color=red,mark=square,mark options={scale=0.7}] file {tradeoff-conad_fairod_reddit.txt};\addlegendentry{\tiny FairOD}        
        \addplot[only marks,color=violet,mark=triangle,mark options={scale=0.7}] file {tradeoff-conad_hin_reddit.txt};\addlegendentry{\tiny HIN}  
        \addplot[only marks,color=blue,solid,mark=oplus,mark options={solid, scale=1},] file {tradeoff-conad_reddit.txt};\addlegendentry{\tiny No Reg.} 
    \end{axis}
    \end{tikzpicture}
\begin{tikzpicture}
    \centering
    \begin{axis}[
        xlabel={AUCROC ($\uparrow$)},
        ylabel={EOO ($\downarrow$)},
        title={\textbf{(b) CONAD on Twitter}},
        height=3cm, width=0.49\linewidth,
        ymin=0.01, ymax=0.06,
        xmin=0.48, xmax=0.57,
        scaled y ticks = false,
        legend columns=2,
        legend pos=north west,
        y tick label style={/pgf/number format/.cd,fixed,fixed zerofill,precision=3,/tikz/.cd},
        ]
        \addplot[only marks, color=teal,mark=o,mark options={scale=0.7}] file {tradeoff-conad_correlation.txt};\addlegendentry{\tiny Corrleation}        
        \addplot[only marks,color=red,mark=square,mark options={scale=0.7}] file {tradeoff-conad_fairod.txt};\addlegendentry{\tiny FairOD}        
        \addplot[only marks,color=violet,mark=triangle,mark options={scale=0.7}] file {tradeoff-conad_hin.txt};\addlegendentry{\tiny HIN}  
        \addplot[only marks,color=blue,solid,mark=oplus,mark options={solid, scale=1},] file {tradeoff-conad.txt};\addlegendentry{\tiny No Reg.} 
    \end{axis}
    \end{tikzpicture}

\begin{tikzpicture}
    \centering
    \begin{axis}[
        xlabel={AUCROC ($\uparrow$)},
        ylabel={EOO ($\downarrow$)},
        title={\textbf{(c) DOMINANT on Reddit}},
        height=3cm, width=0.49\linewidth,
        xtick={0.59, 0.60, 0.61},
        ytick={0.040,0.050,0.060},
        ymin=0.036, ymax=0.061,        
        xmin=0.589, xmax=0.612,
        scaled y ticks = false,
        legend columns=2,
        legend pos=north west,
        y tick label style={/pgf/number format/.cd,fixed,fixed zerofill,precision=3,/tikz/.cd},
        ]              
        \addplot[only marks, color=teal,mark=o,mark options={scale=0.7}] file {tradeoff-dominant_correlation_reddit.txt};\addlegendentry{\tiny Corrleation}        
        \addplot[only marks,color=red,mark=square,mark options={scale=0.7}] file {tradeoff-dominant_fairod_reddit.txt};\addlegendentry{\tiny FairOD}        
        \addplot[only marks,color=violet,mark=triangle,mark options={scale=0.7}] file {tradeoff-dominant_hin_reddit.txt};\addlegendentry{\tiny HIN}  
        \addplot[only marks,color=blue,solid,mark=oplus,mark options={solid, scale=1},] file {tradeoff-dominant_reddit.txt};\addlegendentry{\tiny No Reg.} 
    \end{axis}
    \end{tikzpicture}
\begin{tikzpicture}
    \centering
    \begin{axis}[
        xlabel={AUCROC ($\uparrow$)},
        ylabel={EOO ($\downarrow$)},
         title={\textbf{(d) DOMINANT on Twitter}},
        height=3cm, width=0.49\linewidth,
        ymin=0.01, ymax=0.08,        
        xmin=0.47, xmax=0.62,
        scaled y ticks = false,
        legend columns=2,
        legend pos=north west,
        y tick label style={/pgf/number format/.cd,fixed,fixed zerofill,precision=3,/tikz/.cd},
        ]            
        \addplot[only marks, color=teal,mark=o,mark options={scale=0.7}] file {tradeoff-dominant_correlation.txt};\addlegendentry{\tiny Corrleation}        
        \addplot[only marks,color=red,mark=square,mark options={scale=0.7}] file {tradeoff-dominant_fairod.txt};\addlegendentry{\tiny FairOD}        
        \addplot[only marks,color=violet,mark=triangle,mark options={scale=0.7}] file {tradeoff-dominant_hin.txt};\addlegendentry{\tiny HIN}  
        \addplot[only marks,color=blue,solid,mark=oplus,mark options={solid, scale=1},] file {tradeoff-dominant.txt};\addlegendentry{\tiny No Reg.} 
    \end{axis}
    \end{tikzpicture}

\begin{tikzpicture}
    \centering
    \begin{axis}[
        xlabel={AUCROC ($\uparrow$)},
        ylabel={EOO ($\downarrow$)},
        title={\textbf{(e) VGOD on Reddit}},
        height=3cm, width=0.49\linewidth,
        xtick={0.65, 0.70, 0.75, 0.80},
        ytick={0.30, 0.40,0.50,0.60},
        ymin=0.3, ymax=0.6,        
        xmin=0.68, xmax=0.82,
        scaled y ticks = false,
        legend columns=2,
        legend pos=north west,
        y tick label style={/pgf/number format/.cd,fixed,fixed zerofill,precision=3,/tikz/.cd},
        ]             
        \addplot[only marks, color=teal,mark=o,mark options={scale=0.7}] file {tradeoff-vgod_correlation_reddit.txt};\addlegendentry{\tiny Corrleation}        
        \addplot[only marks,color=red,mark=square,mark options={scale=0.7}] file {tradeoff-vgod_fairod_reddit.txt};\addlegendentry{\tiny FairOD}        
        \addplot[only marks,color=violet,mark=triangle,mark options={scale=0.7}] file {tradeoff-vgod_hin_reddit.txt};\addlegendentry{\tiny HIN}  
        \addplot[only marks,color=blue,solid,mark=oplus,mark options={solid, scale=1},] file {tradeoff-vgod_reddit.txt};\addlegendentry{\tiny No Reg.} 
    \end{axis}
    \end{tikzpicture}
\begin{tikzpicture}
    \centering
    \begin{axis}[
        xlabel={AUCROC ($\uparrow$)},
        ylabel={EOO ($\downarrow$)},
         title={\textbf{(f) VGOD on Twitter}},
        height=3cm, width=0.49\linewidth,
        ymin=0.0, ymax=0.12,        
        xmin=0.7, xmax=0.8,
        scaled y ticks = false,
        legend columns=2,
        legend pos=south east,
        y tick label style={/pgf/number format/.cd,fixed,fixed zerofill,precision=3,/tikz/.cd},
        ]             
        \addplot[only marks, color=teal,mark=o,mark options={scale=0.7}] file {tradeoff-vgod_correlation.txt};\addlegendentry{\tiny Corrleation}        
        \addplot[only marks,color=red,mark=square,mark options={scale=0.7}] file {tradeoff-vgod_fairod.txt};\addlegendentry{\tiny FairOD}        
        \addplot[only marks,color=violet,mark=triangle,mark options={scale=0.7}] file {tradeoff-vgod_hin.txt};\addlegendentry{\tiny HIN}  
        \addplot[only marks,color=blue,solid,mark=oplus,mark options={solid, scale=1},] file {tradeoff-vgod.txt};\addlegendentry{\tiny No Reg.} 
    \end{axis}
    \end{tikzpicture}

\begin{tikzpicture}
    \centering
    \begin{axis}[
        xlabel={AUCROC ($\uparrow$)},
        ylabel={EOO ($\downarrow$)},
        title={\textbf{(g) CoLA on Reddit}},
        height=3cm, width=0.49*\linewidth,
        ymin=0.025, ymax=0.25,
        xmin=0.38, xmax=0.57,
        legend columns=2,
        legend pos=north west,
        y tick label style={/pgf/number format/.cd,fixed,fixed zerofill,precision=3,/tikz/.cd},
        ]
        \addplot[only marks, color=teal,mark=o,mark options={scale=0.7}] file {tradeoff-cola_correlation_reddit.txt};\addlegendentry{\tiny Correlation}        
        \addplot[only marks,color=red,mark=square,mark options={scale=0.7}] file {tradeoff-cola_fairod_reddit.txt};\addlegendentry{\tiny FairOD}        
        \addplot[only marks,color=violet,mark=triangle,mark options={scale=0.7}] file {tradeoff-cola_hin_reddit.txt};\addlegendentry{\tiny HIN}  
        \addplot[only marks,color=blue,solid,mark=oplus,mark options={solid, scale=1},] file {tradeoff-cola_reddit.txt};\addlegendentry{\tiny No Reg.} 
    \end{axis}
    \end{tikzpicture}
\begin{tikzpicture}
    \centering
    \begin{axis}[
        xlabel={AUCROC ($\uparrow$)},
        ylabel={EOO ($\downarrow$)},
        title={\textbf{(h) CoLA on Twitter}},
        height=2.8cm, width=0.49*\linewidth,
        ymin=0, ymax=0.06,
        xmin=0.33, xmax=0.55,
        scaled y ticks = false,
        legend columns=2,
        legend pos=south west,
        y tick label style={/pgf/number format/.cd,fixed,fixed zerofill,precision=3,/tikz/.cd},
        ]
        \addplot[only marks, color=teal,mark=o,mark options={scale=0.7}] file {tradeoff-cola_correlation.txt};\addlegendentry{\tiny Correlation}        
        \addplot[only marks,color=red,mark=square,mark options={scale=0.7}] file {tradeoff-cola_fairod.txt};\addlegendentry{\tiny FairOD}        
        \addplot[only marks,color=violet,mark=triangle,mark options={scale=0.7}] file {tradeoff-cola_hin.txt};\addlegendentry{\tiny HIN}  
        \addplot[only marks,color=blue,solid,mark=oplus,mark options={solid, scale=1},] file {tradeoff-cola.txt};\addlegendentry{\tiny No Reg.} 
    \end{axis}
    \end{tikzpicture}

    \vspace{-0.3cm}
    \caption{Trade-off spaces for GAD methods with fairness regularizers. The ideal FairGAD method should have low EOO and high AUCROC (\ie, the bottom right corner).}\label{fig:tradeoff}
    \vspace{-0.3cm}
\end{figure*}

\vspace{1mm}
\noindent\textbf{Impact of Graph Debiasers (\fairwalk\ and \edits).}
We investigate the impact of debiased graphs and node embeddings obtained through \fairwalk\ and \edits, respectively.
However, we encountered out-of-memory (\ie, `o.o.m') when attempting to obtain the debiased Reddit graph from \edits. 
This is because \edits\ requires the addition of a significant number of new edges (\eg, 97M for Reddit) depending on the average node degree of the input graph.

Interestingly, we observe that the debiased graph from \textbf{\edits} leads to a noticeable improvement in the accuracy of the GAD methods, while their unfairness escalates, as indicated by larger values for SP and EOO (except for \cola).
This observation contradicts the claim made in~\cite{fair_edits} that \edits\ can reduce unfairness while maintaining accuracy.
We speculate that this discrepancy arises from the attribute debiaser used in \edits, which focuses on minimizing the difference in node attribute distributions as a whole, not just the node distributions with respect to the sensitive attribute. 
Furthermore, we observed that \edits\ significantly enhance the accuracy of \conad, \dominant, and \vgod, which fully exploits the augmented graph structure based on the reconstruction error.
However, the accuracy of \cola\ achieved only a minor improvement since it partially exploits the augmented graph structure by sampling node pairs through random walks. 

On the other hand, the modifications made by \textbf{\fairwalk} consistently improve fairness, demonstrated by the decrease in both SP and EOO. 
In terms of accuracy, the GAD methods show different trends whether they use the reconstruction error in the attribute matrix.
Specifically, the accuracy of \dominant, \conad, and \vgod, which jointly learn the reconstruction errors in the adjacency and attribute matrices, decreases, while the accuracy of \cola, which solely relies on the graph structure, increases. 
As mentioned in Section~\ref{sec:setting}, we use the \fairwalk\ embeddings instead of node attributes as node features to reduce the attribute bias.
Thus, the optimization of \dominant, \conad, and \vgod\ becomes more challenging without the use of node attributes.

\vspace{1mm}
\noindent\textbf{Impact of Fairness Regularizers (\fairod, \hin, and \correlation).}
To assess the fairness regularizers, we incorporate them into the original loss function of each GAD method.
Since different regularizers require different weight scales (\ie, $\lambda$ and $\gamma$), we perform separate hyperparameter grid searches for each regularizer. 

\textit{FairOD and HIN}. We investigate how AUCROC (\ie, performance) and EOO (\ie, fairness) vary with changes in the weight of $\lambda$ for $\mathcal{L}_{FairOD}, \mathcal{L}_{HIN}$ and $\gamma$ for $\mathcal{L}_{ADCG}$. 
Due to space limitations, we here report the results of \conad\ since we confirmed that the results of other methods are consistent with those of \conad. 
\autoref{fig:fairod_hin} illustrates the results of \conad\ using \hin\ and \fairod\ with varying $\lambda$ and $\gamma$ values on Reddit.
Regarding the \textbf{performance} metric, where a higher value is better, we observe that increasing $\lambda$ continuously leads to a decrease in AUCROC, and the magnitude of the AUCROC decrease increases as $\lambda$ increases. 
On the other hand, smaller $\gamma$ values result in improved performance, while larger $\gamma$ values lead to decreased performance as \conad\ fails to minimize its original representation loss. 
Regarding the \textbf{fairness} metric, where a lower value is better, increasing both $\lambda$ and $\gamma$ leads to a decrease in EOO as expected from the regularisation term. 
In this context, we note that increasing $\gamma$ can rapidly decrease EOO compared to $\lambda$, as the ADCG loss further encourages minimizing the difference in (true positive) anomaly detection rate between the sensitive attribute groups, rather than simply predicting anomalies at the same rates between them. 
Therefore, the results indicate that we can achieve improvements in both performance and fairness by appropriately setting the values of $\gamma$. 
However, we believe that the gain of improvement is not substantial in either metric.

\textit{Correlation}. 
The impact of \correlation\ is depicted in Figure~\ref{fig:correlation}. 
Since \correlation\
only requires a single weight parameter $\lambda$, we present the results of three GAD methods (\ie, \conad, \dominant, and \vgod) using \correlation\ in \autoref{fig:correlation};
note that the results for \cola\ were removed due to their unreasonably high (for EOO) and low (AUCROC) values.
Except for \cola, the results show that increasing $\lambda$ consistently leads to lower EOO, indicating improved fairness.
However, the magnitude of the performance drop by increasing $\lambda$ varies across methods.
We note 
the differences between the original losses of GAD methods; 
For instance, \dominant\ simply uses the node reconstruction error to rank anomalies, while \conad\ is trained on augmented graphs that encode known anomaly types in addition to the node reconstruction error. 
As such, modifying the joint loss of \conad\ would have a more significant impact on its learning, resulting in decreased performance. 

\textit{CoLA with Fairness Regularizers}. While the fairness regularizers consider reconstruction errors in their formulations, \cola\ relies on the differences between positive and negative neighbor pairs. 
For this reason, the fairness regularizers do not directly contribute to the learning mechanism in \cola. 
When fairness regularizers are introduced, the results of \cola\ exhibit a significant standard deviation, since only emphasizing the losses of certain nodes may not always result in improved performance or fairness.  
This observation highlights the need to develop alternative fairness regularizers that can be effectively incorporated into GAD models with different mechanisms other than reconstruction error.

\textit{Accuracy-Fairness Trade-off Space}. We present the trade-off space between accuracy and fairness for all GAD methods
with the fairness regularizers in \autoref{fig:tradeoff}.
It should be noted that an ideal \ours\ method should achieve high AUCROC and low EOO, which would position it in the bottom right corner in \autoref{fig:tradeoff}.
However, most GAD methods, even after applying the fairness methods, lie along a straight line, indicating a linear trade-off between performance and fairness. 

The trade-off space under the \fairod\ regularizer appears to be worse than that under the \hin\ and \correlation\ regularizers.
For \fairod, the space exhibits a tendency to deviate considerably from the optimal placement in the bottom right corner and displays a significant distance between instances.
This could be attributed to its formulation, which includes the sum and standard deviation of reconstruction errors without a direct link to the sensitive attribute.
On the other hand, \hin\ and \correlation\ 
penalize the method for having a large difference in errors between sensitive attribute groups. 
Given that most GAD methods rely on reconstruction errors to detect anomalies, the formulation of \hin\ and \correlation\ helps to improve the trade-off space to some extent. 

However, none of the existing GAD methods achieve the desired outcomes (\ie, bottom right corner).
This means that it is currently difficult to detect misinformation among right-leaning users. 
As a result, political bias in FairGAD can lead to problems such as reinforcement of confirmation bias, unequal treatment of news sources, and difficulties in achieving fair and accurate categorization.

\begin{table}[t]
\centering
\caption{Changes in EOO ($\downarrow$) across different sensitive attributes on Twitter}
\vspace{-0.25cm}
\resizebox{0.9\linewidth}{!}{
\begin{tabular}{@{}cccc@{}}
\toprule
\textbf{Sensitive Attribute} & \textbf{Political Leaning} & \textbf{Gender}      & \textbf{Age}         \\ \midrule
\textbf{CoLA}                & 0.023±0.012       & 0.007±0.004 & 0.017±0.006 \\
\textbf{CONAD}               & 0.044±0.003       & 0.030±0.004 & 0.036±0.007 \\
\textbf{DOMINANT}            & 0.044±0.003       & 0.031±0.004 & 0.037±0.007 \\
\textbf{VGOD}                & 0.111±0.021       & 0.110±0.033 & 0.055±0.015 \\ \bottomrule
\end{tabular}
\label{tab:gender_results}
}
\vspace{-0.35cm}
\end{table}

\subsection{Sensitive Attribute Analysis}\label{sec:alter_sensitive}

\vspace{1mm}
\noindent\textbf{Alternative Sensitive Attributes.}
We created two new versions of our Twitter dataset, one with \textit{gender} as the sensitive attribute and one with \textit{age} as the sensitive attribute, all of which were inferred by the M3 system.
These attributes are commonly used as sensitive attributes in fairness research (see Table~\ref{tab:statistics}).
Running \cola, \conad, \dominant, and \vgod\ without employing fairness methods on these datasets allowed us to analyze changes in accuracy and fairness metrics according to different sensitive attributes.
It should be noted that the experiments were exclusively performed only on the Twitter dataset due to the incompatibility of the Reddit dataset with the requirements of the M3 system (\ie, general lack of profile images, biographies, and names for Reddit accounts).

\autoref{tab:gender_results} illustrates how the scores of the fairness metric (EOO) vary with sensitive attributes, highlighting different levels of unfairness across attributes.
Selecting political leanings as the sensitive attribute increased levels of unfairness in the Twitter dataset, simplifying the analysis of fairness metric differences after applying fairness regularizers or graph debiasers. 
Here, note that accuracy, as measured by AUCROC, is not reported as it remains relatively constant because the node attributes and the network structure remain unchanged throughout the shifts in the sensitive attribute.

\vspace{1mm}
\noindent\textbf{Elimination of Sensitive Attributes.}
We conducted experiments removing the sensitive attribute from the Reddit dataset. 
Table~\ref{tab:removing_sens} shows the results, where we find a little change in AUCROC and EOO for most GAD methods such as CONAD, DOMINANT, and VGOD that achieve high accuracy.
For COLA, we suspect that the sensitive attribute serves as a significant contrast indicator due to its utilization of contrastive learning between neighbors. 
These results suggest that eliminating the sensitive attribute alone may not sufficiently address the \ours\ problem, as the correlation between the sensitive attribute and the anomaly label can potentially leak into the graph structure and non-sensitive attributes.
This leakage has been demonstrated well in previous research~\citep{fair_edits,WangZDCLD22}.

\begin{table}[t]
\centering
\caption{Changes in AUCROC and EOO according to the elimination of sensitive attributes on Reddit}
\vspace{-0.2cm}
\resizebox{0.87\linewidth}{!}{
\begin{tabular}{@{}ccccc@{}}
\toprule
\textbf{Option} & \multicolumn{2}{c}{\textbf{w/ Sensitive Attributes}}  & \multicolumn{2}{c}{\textbf{w/o Sensitive Attributes}}  \\
\textbf{Metric}  & \textbf{AUCROC ($\uparrow$)}      & \textbf{EOO ($\downarrow$)}         & \textbf{AUCROC ($\uparrow$)}       & \textbf{EOO ($\downarrow$)}       \\ \midrule
\textbf{CoLA}     & 0.453±0.014 & 0.177±0.014 & 0.450±0.014 & 0.035±0.024        \\
\textbf{CONAD}    & 0.608±0.001 & 0.055±0.002 & 0.609±0.001 & 0.056±0.003        \\
\textbf{DOMINANT} & 0.608±0.001 & 0.057±0.003 & 0.608±0.001 & 0.058±0.003       \\ 
\textbf{VGOD} & 0.721±0.009 & 0.472±0.063 & 0.721±0.010 & 0.471±0.064      \\ 
\bottomrule
\end{tabular}
}
\vspace{-0.55cm}
\label{tab:removing_sens}
\end{table}

\section{Ethics Statement}\label{sec:ethics}

\noindent\textbf{User Privacy.}
Our datasets for release are created only from public data available from the Pushshift dataset (refer to \url{https://github.com/pushshift/api}) for Reddit and \citet{verma2022examining} for Twitter. 
We prioritize user privacy by excluding any identifiable information such as usernames or IDs. Each user's postings are represented as low-dimensional embedding vectors, ensuring that raw text is not included.
Therefore, we cannot identify specific users from our datasets nor can we infer an actual user’s political leanings. 
In releasing the data, we adhere to established procedures in data mining and social science research~~\citep{
AoWLQ0020,BeelXSY22,jin2023predicting,jin2024mm,zhao2023competeai,kumar2018community,SaveskiBMR22,xiao2023large} to safeguard user privacy. 
We will provide a Data Use Agreement for researchers accessing the datasets and establish a contact point for users to query their inclusion and request removal of their information.

\vspace{1mm}
\noindent\textbf{Risk of Perpetuating Biases.}
The intention behind this study is NOT to suggest a direct link between political leanings and misinformation propagation, NOR to perpetuate any stereotypes or biases that might result from such a link. 
Instead, our study has two fundamental objectives.
One is to examine whether this correlation actually exists in our datasets, which include real-world user behaviors collected from globally-prominent social media platforms. 
Another is to investigate whether existing GAD methods yield biased outcomes on our datasets due to these inherent biases, and whether existing fairness methods can be effectively incorporated into GAD methods to produce fairer results. 

\vspace{-2mm}
\section{Conclusion}\label{sec:conclusion}

In this work, we defined an important yet under-explored problem, namely \ours, and presented two novel \ours\ datasets that cover aspects of the graph, anomaly labels, and sensitive attributes.
Through extensive experiments, we 
demonstrated that incorporating existing fairness methods into existing GAD methods does not yield the desired outcomes.
This finding emphasizes the need for further investigations and follow-up studies on \ours.

\noindent\textbf{Limitations and Future Work.} 
We defined the \ours\ problem as an unsupervised node classification task
for fair comparison.
Further studies can investigate how semi-supervised learning~\citep{ liuMulGADSemisupervisedGraph2022,wangSemiSupervisedGraphAttentive2019} affects the accuracy and fairness of GAD methods.
Another potential avenue to improve the accuracy-fairness trade-off is by combining  graph debiasers and fairness regularizers. 
Further analysis can investigate the impact and applicability of such combinations across various GAD methods. 

\section*{Acknowledgment}
The work of Srijan Kumar is supported in part by NSF grants CNS-2154118, IIS-2027689, ITE-2137724, ITE-2230692, CNS- 2239879, Defense Advanced Research Projects Agency (DARPA) under Agreement No. HR00112290102 (subcontract No. PO70745), and funding from Microsoft, Google, and The Home Depot.
The work of Yeon-Chang Lee was supported by Institute of Information \& communications Technology Planning \& Evaluation(IITP) grant (No.RS-2020-II201336, Artificial Intelligence graduate school support(UNIST)) and under the Leading Generative AI Human Resources Development (IITP-2024-RS-2024-00360227), both funded by the Korea government(MSIT).
Sang-Wook Kim's work has been supported by the Institute of Information \& Communications Technology Planning \& Evaluation (IITP) grant funded by the Korea government (MSIT) (No.RS-2022-00155586, No.2022-0-00352, No.RS-2020-II201373).

\appendix

\renewcommand\thetable{\Roman{table}}
\renewcommand\thefigure{\Roman{figure}}
\setcounter{table}{0}
\setcounter{figure}{0}

\vspace{+0.2cm}
\begin{center}
\textbf{\LARGE Appendix}    
\end{center}
\section{List of Politics Related Subreddits}\label{sec:pol_subreddits}
We used a crowd-sourced collection\footnote{\url{https://www.reddit.com/r/redditlists/comments/josdr/list_of_political_subreddits/}} of political subreddits~\citep{abs-1711-05303}.

\footnotesize
\noindent``\textsf{r/politics}", ``\textsf{r/Liberal}", ``\textsf{r/Conservative}", ``\textsf{r/Anarchism}",
 ``\textsf{r/LateStageCapitalism}", 
 
 \noindent``\textsf{r/PoliticalDiscussion}",
 ``\textsf{r/PoliticalHumor}", ``\textsf{r/worldpolitics}",
 ``\textsf{r/PoliticalCompassMemes}", ``\textsf{r/PoliticalVideo}",
 ``\textsf{r/PoliticalDiscourse}", ``\textsf{r/PoliticalFactChecking}",
 ``\textsf{r/PoliticalRevisioni-} \textsf{sm}", ``\textsf{r/PoliticalIdeology}",
 ``\textsf{r/PoliticalRevolution}", ``\textsf{r/PoliticalMemes}",
 ``\textsf{r/PoliticalModer-} \textsf{ation}", ``\textsf{r/PoliticalCorrectness}",
 ``\textsf{r/PoliticalCorrectnessGoneMad}", ``\textsf{r/PoliticalTheory}",
 
 \noindent``\textsf{r/PoliticalQuestions}", ``\textsf{r/PoliticalScience}",
 ``\textsf{r/PoliticalHumorModerated}",
 ``\textsf{r/Political} \textsf{Compass}",
 ``\textsf{r/PoliticalDiscussionModerated}",
 ``\textsf{r/worldnews}", ``\textsf{r/news}", ``\textsf{r/worldpolitics}",
 ``\textsf{r/worldevents}", ``\textsf{r/business}", ``\textsf{r/economics}",
 ``\textsf{r/environment}", ``\textsf{r/energy}", ``\textsf{r/law}", 
 
 \noindent``\textsf{r/education}",
 ``\textsf{r/history}", ``\textsf{r/PoliticsPDFs}", ``\textsf{r/WikiLeaks}", ``\textsf{r/SOPA}",
 ``\textsf{r/NewsPorn}", ``\textsf{r/worldnews2}", ``\textsf{r/AnarchistNews}",
 ``\textsf{r/republicofpolitics}", ``\textsf{r/LGBTnews}", ``\textsf{r/politics2}",
 ``\textsf{r/economic2}", ``\textsf{r/environment2}", ``\textsf{r/uspolitics}",
 ``\textsf{r/AmericanPolitics}", ``\textsf{r/AmericanGover-} \textsf{nment}",
 ``\textsf{r/ukpolitics}", ``\textsf{r/canada}", ``\textsf{r/euro}", ``\textsf{r/Palestine}",
 ``\textsf{r/eupolitics}", ``\textsf{r/MiddleEast-} \textsf{News}", ``\textsf{r/Israel}", ``\textsf{r/india}",
 ``\textsf{r/pakistan}", ``\textsf{r/china}", ``\textsf{r/taiwan}", ``\textsf{r/iran}", ``\textsf{r/russia}",
 
 \noindent``\textsf{r/Libertarian}", ``\textsf{r/Anarchism}", ``\textsf{r/socialism}",
 ``\textsf{r/progressive}", ``\textsf{r/Conservative}",
 ``\textsf{r/ameri-} \textsf{canpirateparty}", ``\textsf{r/democrats}", ``\textsf{r/Liberal}",
 ``\textsf{r/new\_right}", ``\textsf{r/Republican}", ``\textsf{r/egalitarian}",
 ``\textsf{r/demsocialist}", ``\textsf{r/LibertarianLeft}", ``\textsf{r/Liberty}",
 ``\textsf{r/Anarcho\_Capitalism}", ``\textsf{r/alltheleft}", ``\textsf{r/neoprogs}",
 ``\textsf{r/democracy}", ``\textsf{r/peoplesparty}", ``\textsf{r/Capitalism}",
 ``\textsf{r/Anarchist}", ``\textsf{r/feminisms}", ``\textsf{r/republicans}",
 ``\textsf{r/Egalitarianism}", ``\textsf{r/anarchafeminism}", ``\textsf{r/Communist}",
 ``\textsf{r/social-} \textsf{democracy}", ``\textsf{r/conservatives}", ``\textsf{r/Freethought}",
 ``\textsf{r/StateOfTheUnion}", ``\textsf{r/equality}", ``\textsf{r/propagandaposters}",
 ``\textsf{r/SocialScience}", ``\textsf{r/racism}", ``\textsf{r/corruption}",
 ``\textsf{r/propaganda}", ``\textsf{r/lgbt}", ``\textsf{r/feminism}", ``\textsf{r/censorship}",
 ``\textsf{r/obama}", ``\textsf{r/war}", ``\textsf{r/antiwar}", ``\textsf{r/climateskeptics}",
 ``\textsf{r/conspiracyhub}", ``\textsf{r/infograffiti}", ``\textsf{r/CalPolitics}", ``\textsf{r/politics\_new}"

\section{Results on Additional Baselines}\label{app:additional_baselines}

\normalsize
\autoref{tab:adone_reddit} demonstrates that the new baselines, including five non-GNN-based anomaly detection methods (\ie, DONE~\citep{BandyopadhyayNV20}, AdONE~\citep{BandyopadhyayNV20}, ECOD~\citep{Li22ecod}, VAE~\citep{KingmaW13}, and ONE~\citep{BandyopadhyayLM19}) and two heuristic methods (\ie, LOF~\citep{BreunigKNS00} and IF~\citep{LiuTZ08}), achieve accuracy levels intermediate to those of the CoLA method and other GNN-based methods.

\begin{table}[t]
\centering
\caption{AUCROC and EOO results across different anomaly detection methods on our datasets}
\vspace{-0.2cm}
\resizebox{0.98\linewidth}{!}{
\begin{tabular}{@{}c|cc|cc|cc@{}}
\toprule
\textbf{Debiasers} & \multicolumn{2}{c|}{\textbf{\boldsymbol{$\times$}}}  & \multicolumn{2}{c|}{\textbf{EDITS}} & \multicolumn{2}{c}{\textbf{FairWalk}} \\
\textbf{Metric}   & \textbf{AUCROC}      & \textbf{EOO}         & \textbf{AUCROC}      & \textbf{EOO}         & \textbf{AUCROC}        & \textbf{EOO}          \\ \midrule
\multicolumn{7}{l}{\textbf{(a) Twitter Dataset}}   \\   \midrule
\textbf{DONE}     & 0.507±0.023 & 0.025±0.015 & 0.577±0.031 & 0.088±0.028 & 0.590±0.014   & 0.079±0.012  \\
\textbf{AdONE}    & 0.522±0.026 & 0.023±0.010 & 0.578±0.032 & 0.101±0.033 & 0.594±0.014   & 0.085±0.013  \\ 
\textbf{ECOD} & 0.454±0.000 & 0.018±0.000 & 0.454±0.000 & 0.018±0.000& 0.704±0.000& 0.157±0.000\\
\textbf{VAE}  & 0.456±0.000 & 0.019±0.000 & 0.457±0.000&0.019±0.000&0.708±0.000&0.158±0.000 \\ 
\textbf{ONE}  & 0.501±0.005 & 0.010±0.008 & 0.501±0.005&0.010±0.008&0.544±0.005&0.025±0.011 \\ \midrule
\textbf{LoF} & 0.460±0.000 & 0.029±0.000 & 0.451±0.000&0.035±0.000&0.500±0.000&0.010±0.000\\
\textbf{IF} & 0.461±0.003 & 0.015±0.005 & 0.461±0.010&0.018±0.001&0.699±0.002&0.145±0.014\\
\bottomrule
\multicolumn{7}{l}{\vspace{-0.2cm}}    \\
\multicolumn{7}{l}{\textbf{(b) Reddit Dataset}}     \\   \midrule                                           
\textbf{DONE}     & 0.578±0.033 & 0.068±0.043 & o.o.m        & o.o.m      & 0.600±0.011   & 0.148±0.015  \\
\textbf{AdONE}    & 0.575±0.027 & 0.077±0.048 & o.o.m        & o.o.m      & 0.607±0.011   & 0.157±0.015  \\ 
\textbf{ECOD} & 0.578±0.000 & 0.098±0.000 & o.o.m&o.o.m&0.736±0.000&0.467±0.000 \\
\textbf{VAE}  & 0.580±0.000 & 0.098±0.000 & o.o.m&o.o.m&0.735±0.000&0.474±0.000 \\ 
\textbf{ONE}   & 0.496±0.007 & 0.014±0.009 & o.o.m&o.o.m&0.524±0.008&0.035±0.021 \\\midrule
\textbf{LoF} & 0.597±0.000 & 0.088±0.000 & o.o.m&o.o.m&0.614±0.000&0.162±0.000\\
\textbf{IF} & 0.580±0.003 & 0.095±0.007 & o.o.m&o.o.m&0.725±0.008&0.428±0.019 \\
\bottomrule
\end{tabular}
}
\vspace{-0.4cm}
\label{tab:adone_reddit}
\end{table}

\section{Further Details on Fairness Regularizers}\label{sec:fairness_equation}

 \normalsize
\textbf{FairOD} \citep{fair_fairod} 
introduces two losses $\mathcal{L}_{FairOD}$ and $\mathcal{L}_{ADCG}$.
The SP loss $\mathcal{L}_{FairOD}$ is formulated as~\citep{fair_fairod}:

\small
\begin{equation}
\mathcal{L}_{FairOD} = \left|\left(1-\frac{1}{n}\right)^2\frac{\left(\sum^{n}_{i=1}Err(v_i)\right)\left(\sum^{n}_{i=1}S(v_i)\right)}{\sigma_{Err} \sigma_S}\right|,
\end{equation}
\normalsize
where $Err(v_i)$ and $S(v_i)$ represent the reconstruction error and sensitive attribute of node $v_i$, respectively.
Also, $\sigma_{Err}$ and $\sigma_S$ represent the standard deviations of the reconstruction error and sensitive attribute, respectively, across all nodes. 
The additional loss $\mathcal{L}_{ADCG}$ is defined as~\citep{fair_fairod}:

\small
\begin{equation}
\mathcal{L}_{ADCG} = \sum_{s\in\{0,1\}}\left(1 - \sum_{\{v_i: S(v_i) = s\}}\frac{2^{BaseErr(v_i)} - 1}{\log_2\left(1 + IDCG_{S=s}\cdot DIFF(v_i)\right)}\right),    
\end{equation}
\normalsize
where $BaseErr(v_i)$ indicates the reconstruction error of node $v_i$ in the original model.
$DIFF(v_i) = \sum_{\{v_j: S(v_j) = s\}}sigmoid$ $(Err(v_j) - Err(v_i))$ represents the differentiable ranking loss utilizing the sigmoid function, and $IDCG_{S=s} = \sum^{|\{v_j: S(v_j) = s\}|}_{j=1} (2^{BaseErr(v_j)} - 1)/(\log_2(1+j))$ is the ideal discounted cumulative gain.

The \textbf{correlation} regularizer, expressed as \footnotesize$\mathcal{L}_{corr} = \left|\frac{(\mathbf{Err}\cdot \mathbf{S})}{\sqrt{(\mathbf{Err}\cdot \mathbf{Err})(\mathbf{S}\cdot \mathbf{S})}}\right|$,\normalsize\ considers the dot product of the reconstruction error (\ie, $\mathbf{Err}$) and sensitive attribute vectors (\ie, $\mathbf{S}$) across all nodes.
Lastly, \textbf{HIN} \citep{fair_hin} penalizes the difference in prediction rates between sensitive attribute groups for both anomalies and non-anomalies:

\small
\begin{equation}
\mathcal{L}_{HIN} = \sum_{y\in\{0,1\}} \left(\frac{\sum_{\{v: S(v) = 1\}}Pr(\hat{y}_v = y)}{|\{v: S(v) = 1\}|} - \frac{\sum_{\{v: S(v) = 0\}}Pr(\hat{y}_v = y)}{|\{v: S(v) = 0\}|}\right)^2,
\end{equation}
\normalsize
where $Pr(\hat{y}_v = 1)$ indicates the probability that node $v$ is predicted as an anomaly. 
Note that it introduces another function that reduces EOO, but requires labels. 
Thus, as mentioned in Section~\ref{sec:setting}, we used $\mathcal{L}_{ADCG}$ from the \fairod\ regularizer as a replacement. 

\section{Further Implementation Details}\label{sec:implementation}

For \cola, \conad, \dominant, and \vgod, 
We used the default hyperparameters provided by PyGOD or their official documentation. 
Batch sampling was used for larger datasets, such as our Twitter dataset and its debiased versions after running the graph debiasers (\ie, \fairwalk\ and \edits), with a batch size of 16,384.
The \fairwalk\ implementation\footnote{\url{https://github.com/urielsinger/fairwalk}} was used with hyperparameters of hidden dimensions=64, walk length=30, number of walks=200, window size=10, and node batch=4. 
The \edits\ implementation\footnote{\url{https://github.com/yushundong/EDITS}} was used with hyperparameters of epoch=500, and learning rate=0.001. 

\newpage
\balance
\bibliography{bibliography}


\begin{thebibliography}{89}


\ifx \showCODEN    \undefined \def \showCODEN     #1{\unskip}     \fi
\ifx \showDOI      \undefined \def \showDOI       #1{#1}\fi
\ifx \showISBNx    \undefined \def \showISBNx     #1{\unskip}     \fi
\ifx \showISBNxiii \undefined \def \showISBNxiii  #1{\unskip}     \fi
\ifx \showISSN     \undefined \def \showISSN      #1{\unskip}     \fi
\ifx \showLCCN     \undefined \def \showLCCN      #1{\unskip}     \fi
\ifx \shownote     \undefined \def \shownote      #1{#1}          \fi
\ifx \showarticletitle \undefined \def \showarticletitle #1{#1}   \fi
\ifx \showURL      \undefined \def \showURL       {\relax}        \fi
\providecommand\bibfield[2]{#2}
\providecommand\bibinfo[2]{#2}
\providecommand\natexlab[1]{#1}
\providecommand\showeprint[2][]{arXiv:#2}

\bibitem[Agarwal et~al\mbox{.}(2021)]%
        {agarwalUnifiedFrameworkFair2021}
\bibfield{author}{\bibinfo{person}{Chirag Agarwal}, \bibinfo{person}{Himabindu Lakkaraju}, {and} \bibinfo{person}{Marinka Zitnik}.} \bibinfo{year}{2021}\natexlab{}.
\newblock \showarticletitle{Towards a Unified Framework for Fair and Stable Graph Representation Learning}. In \bibinfo{booktitle}{\emph{UAI}}, Vol.~\bibinfo{volume}{161}. \bibinfo{pages}{2114--2124}.
\newblock


\bibitem[Ao et~al\mbox{.}(2021)]%
        {AoWLQ0020}
\bibfield{author}{\bibinfo{person}{Xiang Ao}, \bibinfo{person}{Xiting Wang}, \bibinfo{person}{Ling Luo}, \bibinfo{person}{Ying Qiao}, \bibinfo{person}{Qing He}, {and} \bibinfo{person}{Xing Xie}.} \bibinfo{year}{2021}\natexlab{}.
\newblock \showarticletitle{{PENS:} {A} Dataset and Generic Framework for Personalized News Headline Generation}. In \bibinfo{booktitle}{\emph{ACL}}. \bibinfo{pages}{82--92}.
\newblock


\bibitem[Bandyopadhyay et~al\mbox{.}(2019)]%
        {BandyopadhyayLM19}
\bibfield{author}{\bibinfo{person}{Sambaran Bandyopadhyay}, \bibinfo{person}{N. Lokesh}, {and} \bibinfo{person}{M.~Narasimha Murty}.} \bibinfo{year}{2019}\natexlab{}.
\newblock \showarticletitle{Outlier Aware Network Embedding for Attributed Networks}. In \bibinfo{booktitle}{\emph{AAAI}}. \bibinfo{pages}{12--19}.
\newblock


\bibitem[Bandyopadhyay et~al\mbox{.}(2020)]%
        {BandyopadhyayNV20}
\bibfield{author}{\bibinfo{person}{Sambaran Bandyopadhyay}, \bibinfo{person}{Lokesh N}, \bibinfo{person}{Saley~Vishal Vivek}, {and} \bibinfo{person}{M.~Narasimha Murty}.} \bibinfo{year}{2020}\natexlab{}.
\newblock \showarticletitle{Outlier Resistant Unsupervised Deep Architectures for Attributed Network Embedding}. In \bibinfo{booktitle}{\emph{WSDM}}. \bibinfo{pages}{25--33}.
\newblock


\bibitem[Beel et~al\mbox{.}(2022)]%
        {BeelXSY22}
\bibfield{author}{\bibinfo{person}{Jacob Beel}, \bibinfo{person}{Tong Xiang}, \bibinfo{person}{Sandeep Soni}, {and} \bibinfo{person}{Diyi Yang}.} \bibinfo{year}{2022}\natexlab{}.
\newblock \showarticletitle{Linguistic Characterization of Divisive Topics Online: Case Studies on Contentiousness in Abortion, Climate Change, and Gun Control}. In \bibinfo{booktitle}{\emph{ICWSM}}. \bibinfo{pages}{32--42}.
\newblock


\bibitem[Beutel et~al\mbox{.}(2017)]%
        {BeutelCZC17}
\bibfield{author}{\bibinfo{person}{Alex Beutel}, \bibinfo{person}{Jilin Chen}, \bibinfo{person}{Zhe Zhao}, {and} \bibinfo{person}{Ed~H. Chi}.} \bibinfo{year}{2017}\natexlab{}.
\newblock \showarticletitle{Data Decisions and Theoretical Implications when Adversarially Learning Fair Representations}.
\newblock \bibinfo{journal}{\emph{CoRR}}  \bibinfo{volume}{abs/1707.00075} (\bibinfo{year}{2017}).
\newblock


\bibitem[Bin~Naeem and Kamel~Boulos(2021)]%
        {bin2021covid}
\bibfield{author}{\bibinfo{person}{Salman Bin~Naeem} {and} \bibinfo{person}{Maged~N Kamel~Boulos}.} \bibinfo{year}{2021}\natexlab{}.
\newblock \showarticletitle{COVID-19 misinformation online and health literacy: a brief overview}.
\newblock \bibinfo{journal}{\emph{International journal of environmental research and public health}} \bibinfo{volume}{18}, \bibinfo{number}{15} (\bibinfo{year}{2021}), \bibinfo{pages}{8091}.
\newblock


\bibitem[Breunig et~al\mbox{.}(2000)]%
        {BreunigKNS00}
\bibfield{author}{\bibinfo{person}{Markus~M. Breunig}, \bibinfo{person}{Hans{-}Peter Kriegel}, \bibinfo{person}{Raymond~T. Ng}, {and} \bibinfo{person}{J{\"{o}}rg Sander}.} \bibinfo{year}{2000}\natexlab{}.
\newblock \showarticletitle{{LOF:} Identifying Density-Based Local Outliers}. In \bibinfo{booktitle}{\emph{ACM SIGMOD}}. \bibinfo{pages}{93--104}.
\newblock


\bibitem[Chen et~al\mbox{.}(2022b)]%
        {chen2022combating}
\bibfield{author}{\bibinfo{person}{Canyu Chen}, \bibinfo{person}{Haoran Wang}, \bibinfo{person}{Matthew Shapiro}, \bibinfo{person}{Yunyu Xiao}, \bibinfo{person}{Fei Wang}, {and} \bibinfo{person}{Kai Shu}.} \bibinfo{year}{2022}\natexlab{b}.
\newblock \showarticletitle{Combating Health Misinformation in Social Media: Characterization, Detection, Intervention, and Open Issues}.
\newblock \bibinfo{journal}{\emph{arXiv:2211.05289}} (\bibinfo{year}{2022}).
\newblock


\bibitem[Chen et~al\mbox{.}(2022a)]%
        {ChenQLX22}
\bibfield{author}{\bibinfo{person}{Xu Chen}, \bibinfo{person}{Qiu Qiu}, \bibinfo{person}{Changshan Li}, {and} \bibinfo{person}{Kunqing Xie}.} \bibinfo{year}{2022}\natexlab{a}.
\newblock \showarticletitle{GraphAD: {A} Graph Neural Network for Entity-Wise Multivariate Time-Series Anomaly Detection}. In \bibinfo{booktitle}{\emph{SIGIR}}. \bibinfo{pages}{2297--2302}.
\newblock


\bibitem[Cohen et~al\mbox{.}(2020)]%
        {cohen2020correct}
\bibfield{author}{\bibinfo{person}{Elizabeth~L Cohen}, \bibinfo{person}{Anita Atwell~Seate}, \bibinfo{person}{Stephen~M Kromka}, \bibinfo{person}{Andrew Sutherland}, \bibinfo{person}{Matthew Thomas}, \bibinfo{person}{Karissa Skerda}, {and} \bibinfo{person}{Andrew Nicholson}.} \bibinfo{year}{2020}\natexlab{}.
\newblock \showarticletitle{To correct or not to correct? Social identity threats increase willingness to denounce fake news through presumed media influence and hostile media perceptions}.
\newblock \bibinfo{journal}{\emph{Communication Research Reports}} \bibinfo{volume}{37}, \bibinfo{number}{5} (\bibinfo{year}{2020}), \bibinfo{pages}{263--275}.
\newblock


\bibitem[Dai and Wang(2021)]%
        {DaiW21}
\bibfield{author}{\bibinfo{person}{Enyan Dai} {and} \bibinfo{person}{Suhang Wang}.} \bibinfo{year}{2021}\natexlab{}.
\newblock \showarticletitle{Say No to the Discrimination: Learning Fair Graph Neural Networks with Limited Sensitive Attribute Information}. In \bibinfo{booktitle}{\emph{WSDM}}. \bibinfo{pages}{680--688}.
\newblock


\bibitem[Ding et~al\mbox{.}(2019b)]%
        {GAD_DOMINANT}
\bibfield{author}{\bibinfo{person}{Kaize Ding}, \bibinfo{person}{Jundong Li}, \bibinfo{person}{Rohit Bhanushali}, {and} \bibinfo{person}{Huan Liu}.} \bibinfo{year}{2019}\natexlab{b}.
\newblock \showarticletitle{Deep Anomaly Detection on Attributed Networks}. In \bibinfo{booktitle}{\emph{SIAM SDM}}. \bibinfo{pages}{594--602}.
\newblock


\bibitem[Ding et~al\mbox{.}(2019a)]%
        {ding2019interactive}
\bibfield{author}{\bibinfo{person}{Kaize Ding}, \bibinfo{person}{Jundong Li}, {and} \bibinfo{person}{Huan Liu}.} \bibinfo{year}{2019}\natexlab{a}.
\newblock \showarticletitle{Interactive anomaly detection on attributed networks}. In \bibinfo{booktitle}{\emph{ACM WSDM}}. \bibinfo{pages}{357--365}.
\newblock


\bibitem[Ding et~al\mbox{.}(2021)]%
        {ding2021cross}
\bibfield{author}{\bibinfo{person}{Kaize Ding}, \bibinfo{person}{Kai Shu}, \bibinfo{person}{Xuan Shan}, \bibinfo{person}{Jundong Li}, {and} \bibinfo{person}{Huan Liu}.} \bibinfo{year}{2021}\natexlab{}.
\newblock \showarticletitle{Cross-domain graph anomaly detection}.
\newblock \bibinfo{journal}{\emph{IEEE TNNLS}} \bibinfo{volume}{33}, \bibinfo{number}{6} (\bibinfo{year}{2021}), \bibinfo{pages}{2406--2415}.
\newblock


\bibitem[Dong et~al\mbox{.}(2021)]%
        {dong2021individual}
\bibfield{author}{\bibinfo{person}{Yushun Dong}, \bibinfo{person}{Jian Kang}, \bibinfo{person}{Hanghang Tong}, {and} \bibinfo{person}{Jundong Li}.} \bibinfo{year}{2021}\natexlab{}.
\newblock \showarticletitle{Individual fairness for graph neural networks: A ranking based approach}. In \bibinfo{booktitle}{\emph{ACM KDD}}. \bibinfo{pages}{300--310}.
\newblock


\bibitem[Dong et~al\mbox{.}(2022)]%
        {fair_edits}
\bibfield{author}{\bibinfo{person}{Yushun Dong}, \bibinfo{person}{Ninghao Liu}, \bibinfo{person}{Brian Jalaian}, {and} \bibinfo{person}{Jundong Li}.} \bibinfo{year}{2022}\natexlab{}.
\newblock \showarticletitle{{{EDITS}}: {{Modeling}} and {{Mitigating Data Bias}} for {{Graph Neural Networks}}}. In \bibinfo{booktitle}{\emph{TheWebConf}}. \bibinfo{pages}{1259--1269}.
\newblock


\bibitem[Dong et~al\mbox{.}(2023)]%
        {dong2023fairness}
\bibfield{author}{\bibinfo{person}{Yushun Dong}, \bibinfo{person}{Jing Ma}, \bibinfo{person}{Song Wang}, \bibinfo{person}{Chen Chen}, {and} \bibinfo{person}{Jundong Li}.} \bibinfo{year}{2023}\natexlab{}.
\newblock \showarticletitle{Fairness in graph mining: A survey}.
\newblock \bibinfo{journal}{\emph{IEEE TKDE}} (\bibinfo{year}{2023}).
\newblock


\bibitem[Fleiss et~al\mbox{.}(2013)]%
        {fleiss2013statistical}
\bibfield{author}{\bibinfo{person}{Joseph~L Fleiss}, \bibinfo{person}{Bruce Levin}, {and} \bibinfo{person}{Myunghee~Cho Paik}.} \bibinfo{year}{2013}\natexlab{}.
\newblock \bibinfo{booktitle}{\emph{Statistical methods for rates and proportions}}.
\newblock \bibinfo{publisher}{john wiley \& sons}.
\newblock


\bibitem[Grinberg et~al\mbox{.}(2019)]%
        {grinberg2019fake}
\bibfield{author}{\bibinfo{person}{Nir Grinberg}, \bibinfo{person}{Kenneth Joseph}, \bibinfo{person}{Lisa Friedland}, \bibinfo{person}{Briony Swire-Thompson}, {and} \bibinfo{person}{David Lazer}.} \bibinfo{year}{2019}\natexlab{}.
\newblock \showarticletitle{Fake news on Twitter during the 2016 US presidential election}.
\newblock \bibinfo{journal}{\emph{Science}} \bibinfo{volume}{363}, \bibinfo{number}{6425} (\bibinfo{year}{2019}), \bibinfo{pages}{374--378}.
\newblock


\bibitem[Gupta et~al\mbox{.}(2023)]%
        {GuptaDPMD23}
\bibfield{author}{\bibinfo{person}{Manjul Gupta}, \bibinfo{person}{Denis Dennehy}, \bibinfo{person}{Carlos~M. Parra}, \bibinfo{person}{Matti M{\"{a}}ntym{\"{a}}ki}, {and} \bibinfo{person}{Yogesh~K. Dwivedi}.} \bibinfo{year}{2023}\natexlab{}.
\newblock \showarticletitle{Fake news believability: The effects of political beliefs and espoused cultural values}.
\newblock \bibinfo{journal}{\emph{Inf. Manag.}} \bibinfo{volume}{60}, \bibinfo{number}{2} (\bibinfo{year}{2023}), \bibinfo{pages}{103745}.
\newblock


\bibitem[He et~al\mbox{.}(2021)]%
        {HeZSRYK21}
\bibfield{author}{\bibinfo{person}{Bing He}, \bibinfo{person}{Caleb Ziems}, \bibinfo{person}{Sandeep Soni}, \bibinfo{person}{Naren Ramakrishnan}, \bibinfo{person}{Diyi Yang}, {and} \bibinfo{person}{Srijan Kumar}.} \bibinfo{year}{2021}\natexlab{}.
\newblock \showarticletitle{Racism is a virus: anti-asian hate and counterspeech in social media during the {COVID-19} crisis}. In \bibinfo{booktitle}{\emph{ASONAM}}. \bibinfo{pages}{90--94}.
\newblock


\bibitem[Huang et~al\mbox{.}(2023)]%
        {Huang0Z023}
\bibfield{author}{\bibinfo{person}{Yihong Huang}, \bibinfo{person}{Liping Wang}, \bibinfo{person}{Fan Zhang}, {and} \bibinfo{person}{Xuemin Lin}.} \bibinfo{year}{2023}\natexlab{}.
\newblock \showarticletitle{Unsupervised Graph Outlier Detection: Problem Revisit, New Insight, and Superior Method}. In \bibinfo{booktitle}{\emph{IEEE ICDE}}. \bibinfo{pages}{2565--2578}.
\newblock


\bibitem[Jin et~al\mbox{.}(2023a)]%
        {jin2023code}
\bibfield{author}{\bibinfo{person}{Yiqiao Jin}, \bibinfo{person}{Yunsheng Bai}, \bibinfo{person}{Yanqiao Zhu}, \bibinfo{person}{Yizhou Sun}, {and} \bibinfo{person}{Wei Wang}.} \bibinfo{year}{2023}\natexlab{a}.
\newblock \showarticletitle{Code Recommendation for Open Source Software Developers}. In \bibinfo{booktitle}{\emph{ACM Web Conference}}.
\newblock


\bibitem[Jin et~al\mbox{.}(2024)]%
        {jin2024mm}
\bibfield{author}{\bibinfo{person}{Yiqiao Jin}, \bibinfo{person}{Minje Choi}, \bibinfo{person}{Gaurav Verma}, \bibinfo{person}{Jindong Wang}, {and} \bibinfo{person}{Srijan Kumar}.} \bibinfo{year}{2024}\natexlab{}.
\newblock \showarticletitle{MM-Soc: Benchmarking Multimodal Large Language Models in Social Media Platforms}. In \bibinfo{booktitle}{\emph{ACL}}.
\newblock


\bibitem[Jin et~al\mbox{.}(2023b)]%
        {jin2023predicting}
\bibfield{author}{\bibinfo{person}{Yiqiao Jin}, \bibinfo{person}{Yeon-Chang Lee}, \bibinfo{person}{Kartik Sharma}, \bibinfo{person}{Meng Ye}, \bibinfo{person}{Karan Sikka}, \bibinfo{person}{Ajay Divakaran}, {and} \bibinfo{person}{Srijan Kumar}.} \bibinfo{year}{2023}\natexlab{b}.
\newblock \showarticletitle{Predicting Information Pathways Across Online Communities}. In \bibinfo{booktitle}{\emph{ACM KDD}}.
\newblock


\bibitem[Jin et~al\mbox{.}(2022)]%
        {jin2022towards}
\bibfield{author}{\bibinfo{person}{Yiqiao Jin}, \bibinfo{person}{Xiting Wang}, \bibinfo{person}{Ruichao Yang}, \bibinfo{person}{Yizhou Sun}, \bibinfo{person}{Wei Wang}, \bibinfo{person}{Hao Liao}, {and} \bibinfo{person}{Xing Xie}.} \bibinfo{year}{2022}\natexlab{}.
\newblock \showarticletitle{Towards fine-grained reasoning for fake news detection}. In \bibinfo{booktitle}{\emph{AAAI}}, Vol.~\bibinfo{volume}{36}. \bibinfo{pages}{5746--5754}.
\newblock


\bibitem[Kang et~al\mbox{.}(2022)]%
        {kang2022framework}
\bibfield{author}{\bibinfo{person}{David~Y Kang}, \bibinfo{person}{Woncheol Lee}, \bibinfo{person}{Yeon-Chang Lee}, \bibinfo{person}{Kyungsik Han}, {and} \bibinfo{person}{Sang-Wook Kim}.} \bibinfo{year}{2022}\natexlab{}.
\newblock \showarticletitle{A Framework for Accurate Community Detection on Signed Networks Using Adversarial Learning}.
\newblock \bibinfo{journal}{\emph{TKDE}} (\bibinfo{year}{2022}).
\newblock


\bibitem[Kang and Tong(2021)]%
        {KangT21}
\bibfield{author}{\bibinfo{person}{Jian Kang} {and} \bibinfo{person}{Hanghang Tong}.} \bibinfo{year}{2021}\natexlab{}.
\newblock \showarticletitle{Fair Graph Mining}. In \bibinfo{booktitle}{\emph{ACM CIKM}}. \bibinfo{pages}{4849--4852}.
\newblock


\bibitem[Kang et~al\mbox{.}(2021)]%
        {KangLLHK21}
\bibfield{author}{\bibinfo{person}{Yoonsuk Kang}, \bibinfo{person}{Woncheol Lee}, \bibinfo{person}{Yeon{-}Chang Lee}, \bibinfo{person}{Kyungsik Han}, {and} \bibinfo{person}{Sang{-}Wook Kim}.} \bibinfo{year}{2021}\natexlab{}.
\newblock \showarticletitle{Adversarial Learning of Balanced Triangles for Accurate Community Detection on Signed Networks}. In \bibinfo{booktitle}{\emph{{IEEE} {ICDM}}}. \bibinfo{pages}{1150--1155}.
\newblock


\bibitem[Kim et~al\mbox{.}(2022b)]%
        {KimLSL22}
\bibfield{author}{\bibinfo{person}{Hwan Kim}, \bibinfo{person}{Byung~Suk Lee}, \bibinfo{person}{Won{-}Yong Shin}, {and} \bibinfo{person}{Sungsu Lim}.} \bibinfo{year}{2022}\natexlab{b}.
\newblock \showarticletitle{Graph Anomaly Detection With Graph Neural Networks: Current Status and Challenges}.
\newblock \bibinfo{journal}{\emph{{IEEE} Access}}  \bibinfo{volume}{10} (\bibinfo{year}{2022}), \bibinfo{pages}{111820--111829}.
\newblock


\bibitem[Kim et~al\mbox{.}(2023)]%
        {KimLK23}
\bibfield{author}{\bibinfo{person}{Min{-}Jeong Kim}, \bibinfo{person}{Yeon{-}Chang Lee}, {and} \bibinfo{person}{Sang{-}Wook Kim}.} \bibinfo{year}{2023}\natexlab{}.
\newblock \showarticletitle{TrustSGCN: Learning Trustworthiness on Edge Signs for Effective Signed Graph Convolutional Networks}. In \bibinfo{booktitle}{\emph{ACM SIGIR}}. \bibinfo{pages}{2451--2455}.
\newblock


\bibitem[Kim et~al\mbox{.}(2022a)]%
        {KimLSK22}
\bibfield{author}{\bibinfo{person}{Taeri Kim}, \bibinfo{person}{Yeon{-}Chang Lee}, \bibinfo{person}{Kijung Shin}, {and} \bibinfo{person}{Sang{-}Wook Kim}.} \bibinfo{year}{2022}\natexlab{a}.
\newblock \showarticletitle{{MARIO:} Modality-Aware Attention and Modality-Preserving Decoders for Multimedia Recommendation}. In \bibinfo{booktitle}{\emph{{ACM} CIKM}}. \bibinfo{pages}{993--1002}.
\newblock


\bibitem[Kingma and Welling(2014)]%
        {KingmaW13}
\bibfield{author}{\bibinfo{person}{Diederik~P. Kingma} {and} \bibinfo{person}{Max Welling}.} \bibinfo{year}{2014}\natexlab{}.
\newblock \showarticletitle{Auto-Encoding Variational Bayes}. In \bibinfo{booktitle}{\emph{ICLR}}.
\newblock


\bibitem[Kipf and Welling(2017)]%
        {KipfW17}
\bibfield{author}{\bibinfo{person}{Thomas~N. Kipf} {and} \bibinfo{person}{Max Welling}.} \bibinfo{year}{2017}\natexlab{}.
\newblock \showarticletitle{Semi-Supervised Classification with Graph Convolutional Networks}. In \bibinfo{booktitle}{\emph{ICLR}}.
\newblock


\bibitem[Kong et~al\mbox{.}(2022)]%
        {KongKJ0LPK22}
\bibfield{author}{\bibinfo{person}{Taeyong Kong}, \bibinfo{person}{Taeri Kim}, \bibinfo{person}{Jinsung Jeon}, \bibinfo{person}{Jeongwhan Choi}, \bibinfo{person}{Yeon{-}Chang Lee}, \bibinfo{person}{Noseong Park}, {and} \bibinfo{person}{Sang{-}Wook Kim}.} \bibinfo{year}{2022}\natexlab{}.
\newblock \showarticletitle{Linear, or Non-Linear, That is the Question!}. In \bibinfo{booktitle}{\emph{ACM {WSDM}}}. \bibinfo{pages}{517--525}.
\newblock


\bibitem[Krohn and Weninger(2022)]%
        {krohn2022subreddit}
\bibfield{author}{\bibinfo{person}{Rachel Krohn} {and} \bibinfo{person}{Tim Weninger}.} \bibinfo{year}{2022}\natexlab{}.
\newblock \showarticletitle{Subreddit Links Drive Community Creation and User Engagement on Reddit}. In \bibinfo{booktitle}{\emph{ICWSM}}, Vol.~\bibinfo{volume}{16}. \bibinfo{pages}{536--547}.
\newblock


\bibitem[Kumar et~al\mbox{.}(2018)]%
        {kumar2018community}
\bibfield{author}{\bibinfo{person}{Srijan Kumar}, \bibinfo{person}{William~L Hamilton}, \bibinfo{person}{Jure Leskovec}, {and} \bibinfo{person}{Dan Jurafsky}.} \bibinfo{year}{2018}\natexlab{}.
\newblock \showarticletitle{Community interaction and conflict on the web}. In \bibinfo{booktitle}{\emph{WWW}}. \bibinfo{pages}{933--943}.
\newblock


\bibitem[Kumar et~al\mbox{.}(2019)]%
        {kumar2019predicting}
\bibfield{author}{\bibinfo{person}{Srijan Kumar}, \bibinfo{person}{Xikun Zhang}, {and} \bibinfo{person}{Jure Leskovec}.} \bibinfo{year}{2019}\natexlab{}.
\newblock \showarticletitle{Predicting dynamic embedding trajectory in temporal interaction networks}. In \bibinfo{booktitle}{\emph{ACM KDD}}. \bibinfo{pages}{1269--1278}.
\newblock


\bibitem[Lakha et~al\mbox{.}(2022)]%
        {lakhaAnomalyDetectionCybersecurity2022}
\bibfield{author}{\bibinfo{person}{Bishal Lakha}, \bibinfo{person}{Sara~Lilly Mount}, \bibinfo{person}{Edoardo Serra}, {and} \bibinfo{person}{Alfredo Cuzzocrea}.} \bibinfo{year}{2022}\natexlab{}.
\newblock \showarticletitle{Anomaly {{Detection}} in {{Cybersecurity Events Through Graph Neural Network}} and {{Transformer Based Model}}: {{A Case Study}} with {{BETH Dataset}}}. In \bibinfo{booktitle}{\emph{IEEE BigData}}. \bibinfo{pages}{5756--5764}.
\newblock


\bibitem[Lawson and Kakkar(2022)]%
        {lawson2022pandemics}
\bibfield{author}{\bibinfo{person}{M~Asher Lawson} {and} \bibinfo{person}{Hemant Kakkar}.} \bibinfo{year}{2022}\natexlab{}.
\newblock \showarticletitle{Of pandemics, politics, and personality: The role of conscientiousness and political ideology in the sharing of fake news.}
\newblock \bibinfo{journal}{\emph{Journal of Experimental Psychology: General}} \bibinfo{volume}{151}, \bibinfo{number}{5} (\bibinfo{year}{2022}), \bibinfo{pages}{1154}.
\newblock


\bibitem[Lee et~al\mbox{.}(2021)]%
        {LeeLLK21}
\bibfield{author}{\bibinfo{person}{Wonchang Lee}, \bibinfo{person}{Yeon{-}Chang Lee}, \bibinfo{person}{Dongwon Lee}, {and} \bibinfo{person}{Sang{-}Wook Kim}.} \bibinfo{year}{2021}\natexlab{}.
\newblock \showarticletitle{Look Before You Leap: Confirming Edge Signs in Random Walk with Restart for Personalized Node Ranking in Signed Networks}. In \bibinfo{booktitle}{\emph{ACM {SIGIR}}}. \bibinfo{pages}{143--152}.
\newblock


\bibitem[Lee et~al\mbox{.}(2018)]%
        {LeeK018}
\bibfield{author}{\bibinfo{person}{Yeon{-}Chang Lee}, \bibinfo{person}{Sang{-}Wook Kim}, {and} \bibinfo{person}{Dongwon Lee}.} \bibinfo{year}{2018}\natexlab{}.
\newblock \showarticletitle{gOCCF: Graph-Theoretic One-Class Collaborative Filtering Based on Uninteresting Items}. In \bibinfo{booktitle}{\emph{{AAAI}}}. \bibinfo{pages}{3448--3456}.
\newblock


\bibitem[Lee et~al\mbox{.}(2022)]%
        {LeeL0K22}
\bibfield{author}{\bibinfo{person}{Yeon{-}Chang Lee}, \bibinfo{person}{JaeHyun Lee}, \bibinfo{person}{Dongwon Lee}, {and} \bibinfo{person}{Sang{-}Wook Kim}.} \bibinfo{year}{2022}\natexlab{}.
\newblock \showarticletitle{{THOR:} Self-Supervised Temporal Knowledge Graph Embedding via Three-Tower Graph Convolutional Networks}. In \bibinfo{booktitle}{\emph{{IEEE} {ICDM}}}. \bibinfo{pages}{1035--1040}.
\newblock


\bibitem[Li et~al\mbox{.}(2022)]%
        {Li22ecod}
\bibfield{author}{\bibinfo{person}{Zheng Li}, \bibinfo{person}{Yue Zhao}, \bibinfo{person}{Xiyang Hu}, \bibinfo{person}{Nicola Botta}, \bibinfo{person}{Cezar Ionescu}, {and} \bibinfo{person}{George~H. Chen}.} \bibinfo{year}{2022}\natexlab{}.
\newblock \showarticletitle{{ECOD:} Unsupervised Outlier Detection Using Empirical Cumulative Distribution Functions}.
\newblock \bibinfo{journal}{\emph{CoRR}}  \bibinfo{volume}{abs/2201.00382} (\bibinfo{year}{2022}).
\newblock
\urldef\tempurl%
\url{https://arxiv.org/abs/2201.00382}
\showURL{%
\tempurl}


\bibitem[Liang et~al\mbox{.}(2024)]%
        {liang2024survey}
\bibfield{author}{\bibinfo{person}{Ke Liang}, \bibinfo{person}{Lingyuan Meng}, \bibinfo{person}{Meng Liu}, \bibinfo{person}{Yue Liu}, \bibinfo{person}{Wenxuan Tu}, \bibinfo{person}{Siwei Wang}, \bibinfo{person}{Sihang Zhou}, \bibinfo{person}{Xinwang Liu}, \bibinfo{person}{Fuchun Sun}, {and} \bibinfo{person}{Kunlun He}.} \bibinfo{year}{2024}\natexlab{}.
\newblock \showarticletitle{A survey of knowledge graph reasoning on graph types: Static, dynamic, and multi-modal}.
\newblock \bibinfo{journal}{\emph{IEEE Transactions on Pattern Analysis and Machine Intelligence}} (\bibinfo{year}{2024}).
\newblock


\bibitem[Liu et~al\mbox{.}(2008)]%
        {LiuTZ08}
\bibfield{author}{\bibinfo{person}{Fei~Tony Liu}, \bibinfo{person}{Kai~Ming Ting}, {and} \bibinfo{person}{Zhi{-}Hua Zhou}.} \bibinfo{year}{2008}\natexlab{}.
\newblock \showarticletitle{Isolation Forest}. In \bibinfo{booktitle}{\emph{IEEE ICDM}}. \bibinfo{pages}{413--422}.
\newblock


\bibitem[Liu et~al\mbox{.}(2022b)]%
        {liu2022pygod}
\bibfield{author}{\bibinfo{person}{Kay Liu}, \bibinfo{person}{Yingtong Dou}, \bibinfo{person}{Yue Zhao}, \bibinfo{person}{Xueying Ding}, \bibinfo{person}{Xiyang Hu}, \bibinfo{person}{Ruitong Zhang}, \bibinfo{person}{Kaize Ding}, \bibinfo{person}{Canyu Chen}, \bibinfo{person}{Hao Peng}, \bibinfo{person}{Kai Shu}, \bibinfo{person}{George~H. Chen}, \bibinfo{person}{Zhihao Jia}, {and} \bibinfo{person}{Philip~S. Yu}.} \bibinfo{year}{2022}\natexlab{b}.
\newblock \showarticletitle{{{PyGOD}}: {{A}} Python Library for Graph Outlier Detection}.
\newblock \bibinfo{journal}{\emph{arXiv:2204.12095}} (\bibinfo{year}{2022}).
\newblock


\bibitem[Liu et~al\mbox{.}(2022c)]%
        {liu2022bond}
\bibfield{author}{\bibinfo{person}{Kay Liu}, \bibinfo{person}{Yingtong Dou}, \bibinfo{person}{Yue Zhao}, \bibinfo{person}{Xueying Ding}, \bibinfo{person}{Xiyang Hu}, \bibinfo{person}{Ruitong Zhang}, \bibinfo{person}{Kaize Ding}, \bibinfo{person}{Canyu Chen}, \bibinfo{person}{Hao Peng}, \bibinfo{person}{Kai Shu}, \bibinfo{person}{Lichao Sun}, \bibinfo{person}{Jundong Li}, \bibinfo{person}{George~H. Chen}, \bibinfo{person}{Zhihao Jia}, {and} \bibinfo{person}{Philip~S. Yu}.} \bibinfo{year}{2022}\natexlab{c}.
\newblock \showarticletitle{Bond: {{Benchmarking}} Unsupervised Outlier Node Detection on Static Attributed Graphs}.
\newblock \bibinfo{journal}{\emph{NeurIPS}}  \bibinfo{volume}{35} (\bibinfo{year}{2022}), \bibinfo{pages}{27021--27035}.
\newblock


\bibitem[Liu and Liu(2021)]%
        {MNCI_ML_SIGIR}
\bibfield{author}{\bibinfo{person}{Meng Liu} {and} \bibinfo{person}{Yong Liu}.} \bibinfo{year}{2021}\natexlab{}.
\newblock \showarticletitle{Inductive representation learning in temporal networks via mining neighborhood and community influences}. In \bibinfo{booktitle}{\emph{SIGIR}}. \bibinfo{pages}{2202--2206}.
\newblock


\bibitem[Liu et~al\mbox{.}(2024)]%
        {TGC_ML_ICLR}
\bibfield{author}{\bibinfo{person}{Meng Liu}, \bibinfo{person}{Yue Liu}, \bibinfo{person}{Ke Liang}, \bibinfo{person}{Wenxuan Tu}, \bibinfo{person}{Siwei Wang}, \bibinfo{person}{Sihang Zhou}, {and} \bibinfo{person}{Xinwang Liu}.} \bibinfo{year}{2024}\natexlab{}.
\newblock \showarticletitle{Deep Temporal Graph Clustering}. In \bibinfo{booktitle}{\emph{The 12th International Conference on Learning Representations}}.
\newblock


\bibitem[Liu et~al\mbox{.}(2023)]%
        {TGC_ML}
\bibfield{author}{\bibinfo{person}{Meng Liu}, \bibinfo{person}{Yue Liu}, \bibinfo{person}{Ke Liang}, \bibinfo{person}{Siwei Wang}, \bibinfo{person}{Sihang Zhou}, {and} \bibinfo{person}{Xinwang Liu}.} \bibinfo{year}{2023}\natexlab{}.
\newblock \showarticletitle{Deep Temporal Graph Clustering}.
\newblock \bibinfo{journal}{\emph{arXiv:2305.10738}} (\bibinfo{year}{2023}).
\newblock


\bibitem[Liu et~al\mbox{.}(2021)]%
        {GAD_COLA}
\bibfield{author}{\bibinfo{person}{Yixin Liu}, \bibinfo{person}{Zhao Li}, \bibinfo{person}{Shirui Pan}, \bibinfo{person}{Chen Gong}, \bibinfo{person}{Chuan Zhou}, {and} \bibinfo{person}{George Karypis}.} \bibinfo{year}{2021}\natexlab{}.
\newblock \showarticletitle{Anomaly {{Detection}} on {{Attributed Networks}} via {{Contrastive Self-Supervised Learning}}}.
\newblock \bibinfo{journal}{\emph{IEEE TNNLS}} \bibinfo{volume}{33}, \bibinfo{number}{6} (\bibinfo{year}{2021}), \bibinfo{pages}{2378--2392}.
\newblock


\bibitem[Liu et~al\mbox{.}(2022a)]%
        {liuMulGADSemisupervisedGraph2022}
\bibfield{author}{\bibinfo{person}{Zhiyuan Liu}, \bibinfo{person}{Chunjie Cao}, {and} \bibinfo{person}{Jingzhang Sun}.} \bibinfo{year}{2022}\natexlab{a}.
\newblock \showarticletitle{Mul-{{GAD}}: A Semi-Supervised Graph Anomaly Detection Framework via Aggregating Multi-View Information}.
\newblock \bibinfo{journal}{\emph{arXiv:2212.05478}} (\bibinfo{year}{2022}).
\newblock


\bibitem[Louizos et~al\mbox{.}(2016)]%
        {LouizosSLWZ15}
\bibfield{author}{\bibinfo{person}{Christos Louizos}, \bibinfo{person}{Kevin Swersky}, \bibinfo{person}{Yujia Li}, \bibinfo{person}{Max Welling}, {and} \bibinfo{person}{Richard~S. Zemel}.} \bibinfo{year}{2016}\natexlab{}.
\newblock \showarticletitle{The Variational Fair Autoencoder}. In \bibinfo{booktitle}{\emph{ICLR}}.
\newblock


\bibitem[Lu et~al\mbox{.}(2023)]%
        {LuAKZ23}
\bibfield{author}{\bibinfo{person}{Meichen Lu}, \bibinfo{person}{Maged Ali}, \bibinfo{person}{Niraj Kumar}, {and} \bibinfo{person}{Wen Zhang}.} \bibinfo{year}{2023}\natexlab{}.
\newblock \showarticletitle{Identification of the impact of content-related factors on the diffusion of misinformation: {A} Case study of the government intervention polices during {COVID-19} pandemic in the {UK}}. In \bibinfo{booktitle}{\emph{AMCIS}}.
\newblock


\bibitem[Ma et~al\mbox{.}(2021)]%
        {ma2021survey}
\bibfield{author}{\bibinfo{person}{Xiaoxiao Ma}, \bibinfo{person}{Jia Wu}, \bibinfo{person}{Shan Xue}, \bibinfo{person}{Jian Yang}, \bibinfo{person}{Chuan Zhou}, \bibinfo{person}{Quan~Z. Sheng}, \bibinfo{person}{Hui Xiong}, {and} \bibinfo{person}{Leman Akoglu}.} \bibinfo{year}{2021}\natexlab{}.
\newblock \showarticletitle{A Comprehensive Survey on Graph Anomaly Detection with Deep Learning}.
\newblock \bibinfo{journal}{\emph{IEEE TKDE}} (\bibinfo{year}{2021}).
\newblock


\bibitem[Ma et~al\mbox{.}(2023)]%
        {ma2023characterizing}
\bibfield{author}{\bibinfo{person}{Yingchen Ma}, \bibinfo{person}{Bing He}, \bibinfo{person}{Nathan Subrahmanian}, {and} \bibinfo{person}{Srijan Kumar}.} \bibinfo{year}{2023}\natexlab{}.
\newblock \showarticletitle{Characterizing and Predicting Social Correction on Twitter}. In \bibinfo{booktitle}{\emph{WebSci}}. \bibinfo{pages}{86--95}.
\newblock


\bibitem[Micallef et~al\mbox{.}(2020)]%
        {MicallefHKAM20}
\bibfield{author}{\bibinfo{person}{Nicholas Micallef}, \bibinfo{person}{Bing He}, \bibinfo{person}{Srijan Kumar}, \bibinfo{person}{Mustaque Ahamad}, {and} \bibinfo{person}{Nasir~D. Memon}.} \bibinfo{year}{2020}\natexlab{}.
\newblock \showarticletitle{The Role of the Crowd in Countering Misinformation: {A} Case Study of the {COVID-19} Infodemic}. In \bibinfo{booktitle}{\emph{IEEE BigData}}. \bibinfo{pages}{748--757}.
\newblock


\bibitem[Niepert et~al\mbox{.}(2016)]%
        {NiepertAK16}
\bibfield{author}{\bibinfo{person}{Mathias Niepert}, \bibinfo{person}{Mohamed Ahmed}, {and} \bibinfo{person}{Konstantin Kutzkov}.} \bibinfo{year}{2016}\natexlab{}.
\newblock \showarticletitle{Learning Convolutional Neural Networks for Graphs}. In \bibinfo{booktitle}{\emph{ICML}}, Vol.~\bibinfo{volume}{48}. \bibinfo{pages}{2014--2023}.
\newblock


\bibitem[Nithyanand et~al\mbox{.}(2017)]%
        {abs-1711-05303}
\bibfield{author}{\bibinfo{person}{Rishab Nithyanand}, \bibinfo{person}{Brian Schaffner}, {and} \bibinfo{person}{Phillipa Gill}.} \bibinfo{year}{2017}\natexlab{}.
\newblock \showarticletitle{Online Political Discourse in the Trump Era}.
\newblock \bibinfo{journal}{\emph{CoRR}}  \bibinfo{volume}{abs/1711.05303} (\bibinfo{year}{2017}).
\newblock
\urldef\tempurl%
\url{http://arxiv.org/abs/1711.05303}
\showURL{%
\tempurl}


\bibitem[Rahman et~al\mbox{.}(2019)]%
        {fair_fairwalk}
\bibfield{author}{\bibinfo{person}{Tahleen Rahman}, \bibinfo{person}{Bartlomiej Surma}, \bibinfo{person}{Michael Backes}, {and} \bibinfo{person}{Yang Zhang}.} \bibinfo{year}{2019}\natexlab{}.
\newblock \showarticletitle{Fairwalk: {{Towards Fair Graph Embedding}}}. In \bibinfo{booktitle}{\emph{IJCAI}}.
\newblock


\bibitem[Reimers and Gurevych(2019)]%
        {reimersSentenceBERTSentenceEmbeddings2019}
\bibfield{author}{\bibinfo{person}{Nils Reimers} {and} \bibinfo{person}{Iryna Gurevych}.} \bibinfo{year}{2019}\natexlab{}.
\newblock \showarticletitle{Sentence-{{BERT}}: {{Sentence}} Embeddings Using Siamese {{BERT-Networks}}}. In \bibinfo{booktitle}{\emph{EMNLP}}.
\newblock


\bibitem[Reimers and Gurevych(2020)]%
        {reimersMakingMonolingualSentence2020}
\bibfield{author}{\bibinfo{person}{Nils Reimers} {and} \bibinfo{person}{Iryna Gurevych}.} \bibinfo{year}{2020}\natexlab{}.
\newblock \showarticletitle{Making Monolingual Sentence Embeddings Multilingual Using Knowledge Distillation}. In \bibinfo{booktitle}{\emph{EMNLP}}.
\newblock


\bibitem[Sakketou et~al\mbox{.}(2022)]%
        {sakketouFACTOIDNewDataset2022}
\bibfield{author}{\bibinfo{person}{Flora Sakketou}, \bibinfo{person}{Joan Plepi}, \bibinfo{person}{Riccardo Cervero}, \bibinfo{person}{Henri~Jacques Geiss}, \bibinfo{person}{Paolo Rosso}, {and} \bibinfo{person}{Lucie Flek}.} \bibinfo{year}{2022}\natexlab{}.
\newblock \showarticletitle{{{FACTOID}}: {{A}} New Dataset for Identifying Misinformation Spreaders and Political Bias}. In \bibinfo{booktitle}{\emph{LREC}}. \bibinfo{pages}{3231--3241}.
\newblock


\bibitem[Saveski et~al\mbox{.}(2022)]%
        {SaveskiBMR22}
\bibfield{author}{\bibinfo{person}{Martin Saveski}, \bibinfo{person}{Doug Beeferman}, \bibinfo{person}{David McClure}, {and} \bibinfo{person}{Deb Roy}.} \bibinfo{year}{2022}\natexlab{}.
\newblock \showarticletitle{Engaging Politically Diverse Audiences on Social Media}. In \bibinfo{booktitle}{\emph{ICWSM}}. \bibinfo{pages}{873--884}.
\newblock


\bibitem[Semrush(2023)]%
        {semrushreddit}
\bibfield{author}{\bibinfo{person}{Semrush}.} \bibinfo{year}{2023}\natexlab{}.
\newblock \bibinfo{booktitle}{\emph{Reddit Website Traffic, Ranking, Analytics}}.
\newblock
\urldef\tempurl%
\url{https://www.semrush.com/website/reddit.com/overview/}
\showURL{%
\tempurl}


\bibitem[Sharma et~al\mbox{.}(2024)]%
        {SharmaLNSSKK24}
\bibfield{author}{\bibinfo{person}{Kartik Sharma}, \bibinfo{person}{Yeon{-}Chang Lee}, \bibinfo{person}{Sivagami Nambi}, \bibinfo{person}{Aditya Salian}, \bibinfo{person}{Shlok Shah}, \bibinfo{person}{Sang{-}Wook Kim}, {and} \bibinfo{person}{Srijan Kumar}.} \bibinfo{year}{2024}\natexlab{}.
\newblock \showarticletitle{A Survey of Graph Neural Networks for Social Recommender Systems}.
\newblock \bibinfo{journal}{\emph{{ACM} Comput. Surv.}} \bibinfo{volume}{56}, \bibinfo{number}{10} (\bibinfo{year}{2024}), \bibinfo{pages}{265}.
\newblock


\bibitem[Shekhar et~al\mbox{.}(2021)]%
        {fair_fairod}
\bibfield{author}{\bibinfo{person}{Shubhranshu Shekhar}, \bibinfo{person}{Neil Shah}, {and} \bibinfo{person}{Leman Akoglu}.} \bibinfo{year}{2021}\natexlab{}.
\newblock \showarticletitle{{{FairOD}}: {{Fairness-aware Outlier Detection}}}. In \bibinfo{booktitle}{\emph{AIES}}.
\newblock


\bibitem[Shu et~al\mbox{.}(2019)]%
        {shu2019defend}
\bibfield{author}{\bibinfo{person}{Kai Shu}, \bibinfo{person}{Limeng Cui}, \bibinfo{person}{Suhang Wang}, \bibinfo{person}{Dongwon Lee}, {and} \bibinfo{person}{Huan Liu}.} \bibinfo{year}{2019}\natexlab{}.
\newblock \showarticletitle{defend: Explainable fake news detection}. In \bibinfo{booktitle}{\emph{ACM KDD}}. \bibinfo{pages}{395--405}.
\newblock


\bibitem[Verma et~al\mbox{.}(2022)]%
        {verma2022examining}
\bibfield{author}{\bibinfo{person}{Gaurav Verma}, \bibinfo{person}{Ankur Bhardwaj}, \bibinfo{person}{Talayeh Aledavood}, \bibinfo{person}{Munmun De~Choudhury}, {and} \bibinfo{person}{Srijan Kumar}.} \bibinfo{year}{2022}\natexlab{}.
\newblock \showarticletitle{Examining the impact of sharing COVID-19 misinformation online on mental health}.
\newblock \bibinfo{journal}{\emph{Scientific Reports}} \bibinfo{volume}{12}, \bibinfo{number}{1} (\bibinfo{year}{2022}), \bibinfo{pages}{1--9}.
\newblock


\bibitem[Villani(2021)]%
        {villani2021topics}
\bibfield{author}{\bibinfo{person}{C{\'e}dric Villani}.} \bibinfo{year}{2021}\natexlab{}.
\newblock \bibinfo{booktitle}{\emph{Topics in optimal transportation}}. Vol.~\bibinfo{volume}{58}.
\newblock \bibinfo{publisher}{American Mathematical Soc.}
\newblock


\bibitem[Waller and Anderson(2019)]%
        {waller2019generalists}
\bibfield{author}{\bibinfo{person}{Isaac Waller} {and} \bibinfo{person}{Ashton Anderson}.} \bibinfo{year}{2019}\natexlab{}.
\newblock \showarticletitle{Generalists and specialists: Using community embeddings to quantify activity diversity in online platforms}. In \bibinfo{booktitle}{\emph{TheWebConf}}. \bibinfo{pages}{1954--1964}.
\newblock


\bibitem[Wang et~al\mbox{.}(2019b)]%
        {wangSemiSupervisedGraphAttentive2019}
\bibfield{author}{\bibinfo{person}{Daixin Wang}, \bibinfo{person}{Yuan Qi}, \bibinfo{person}{Jianbin Lin}, \bibinfo{person}{Peng Cui}, \bibinfo{person}{Quanhui Jia}, \bibinfo{person}{Zhen Wang}, \bibinfo{person}{Yanming Fang}, \bibinfo{person}{Quan Yu}, \bibinfo{person}{Jun Zhou}, {and} \bibinfo{person}{Shuang Yang}.} \bibinfo{year}{2019}\natexlab{b}.
\newblock \showarticletitle{A {{Semi-Supervised Graph Attentive Network}} for {{Financial Fraud Detection}}}. In \bibinfo{booktitle}{\emph{IEEE ICDM}}. \bibinfo{pages}{598--607}.
\newblock


\bibitem[Wang et~al\mbox{.}(2023)]%
        {wang2023attacking}
\bibfield{author}{\bibinfo{person}{Haoran Wang}, \bibinfo{person}{Yingtong Dou}, \bibinfo{person}{Canyu Chen}, \bibinfo{person}{Lichao Sun}, \bibinfo{person}{Philip~S Yu}, {and} \bibinfo{person}{Kai Shu}.} \bibinfo{year}{2023}\natexlab{}.
\newblock \showarticletitle{Attacking Fake News Detectors via Manipulating News Social Engagement}. In \bibinfo{booktitle}{\emph{TheWebConf}}. \bibinfo{pages}{3978--3986}.
\newblock


\bibitem[Wang et~al\mbox{.}(2022a)]%
        {wang2022training}
\bibfield{author}{\bibinfo{person}{Haonan Wang}, \bibinfo{person}{Ziwei Wu}, {and} \bibinfo{person}{Jingrui He}.} \bibinfo{year}{2022}\natexlab{a}.
\newblock \showarticletitle{Training fair deep neural networks by balancing influence}.
\newblock \bibinfo{journal}{\emph{arXiv:2201.05759}} (\bibinfo{year}{2022}).
\newblock


\bibitem[Wang et~al\mbox{.}(2022b)]%
        {WangZDCLD22}
\bibfield{author}{\bibinfo{person}{Yu Wang}, \bibinfo{person}{Yuying Zhao}, \bibinfo{person}{Yushun Dong}, \bibinfo{person}{Huiyuan Chen}, \bibinfo{person}{Jundong Li}, {and} \bibinfo{person}{Tyler Derr}.} \bibinfo{year}{2022}\natexlab{b}.
\newblock \showarticletitle{Improving Fairness in Graph Neural Networks via Mitigating Sensitive Attribute Leakage}. In \bibinfo{booktitle}{\emph{ACM KDD}}. \bibinfo{publisher}{{ACM}}, \bibinfo{pages}{1938--1948}.
\newblock


\bibitem[Wang et~al\mbox{.}(2019a)]%
        {wang2019demographic}
\bibfield{author}{\bibinfo{person}{Zijian Wang}, \bibinfo{person}{Scott Hale}, \bibinfo{person}{David~Ifeoluwa Adelani}, \bibinfo{person}{Przemyslaw Grabowicz}, \bibinfo{person}{Timo Hartman}, \bibinfo{person}{Fabian Fl{\"o}ck}, {and} \bibinfo{person}{David Jurgens}.} \bibinfo{year}{2019}\natexlab{a}.
\newblock \showarticletitle{Demographic inference and representative population estimates from multilingual social media data}. In \bibinfo{booktitle}{\emph{WWW}}. \bibinfo{pages}{2056--2067}.
\newblock


\bibitem[Wikipedia(2023)]%
        {wikipediamostvisited}
\bibfield{author}{\bibinfo{person}{Wikipedia}.} \bibinfo{year}{2023}\natexlab{}.
\newblock \bibinfo{booktitle}{\emph{List of Most Visited Websites}}.
\newblock
\urldef\tempurl%
\url{https://en.wikipedia.org/wiki/List_of_most_visited_websites}
\showURL{%
\tempurl}


\bibitem[Wu et~al\mbox{.}(2022)]%
        {wu2022bias}
\bibfield{author}{\bibinfo{person}{Junfei Wu}, \bibinfo{person}{Qiang Liu}, \bibinfo{person}{Weizhi Xu}, {and} \bibinfo{person}{Shu Wu}.} \bibinfo{year}{2022}\natexlab{}.
\newblock \showarticletitle{Bias mitigation for evidence-aware fake news detection by causal intervention}. In \bibinfo{booktitle}{\emph{SIGIR}}. \bibinfo{pages}{2308--2313}.
\newblock


\bibitem[Xiao et~al\mbox{.}(2023)]%
        {xiao2023large}
\bibfield{author}{\bibinfo{person}{Yijia Xiao}, \bibinfo{person}{Yiqiao Jin}, \bibinfo{person}{Yushi Bai}, \bibinfo{person}{Yue Wu}, \bibinfo{person}{Xianjun Yang}, \bibinfo{person}{Xiao Luo}, \bibinfo{person}{Wenchao Yu}, \bibinfo{person}{Xujiang Zhao}, \bibinfo{person}{Yanchi Liu}, \bibinfo{person}{Haifeng Chen}, {et~al\mbox{.}}} \bibinfo{year}{2023}\natexlab{}.
\newblock \showarticletitle{Large Language Models Can Be Good Privacy Protection Learners}.
\newblock \bibinfo{journal}{\emph{arXiv:2310.02469}} (\bibinfo{year}{2023}).
\newblock


\bibitem[Xu et~al\mbox{.}(2022)]%
        {xu2022contrastive}
\bibfield{author}{\bibinfo{person}{Zhiming Xu}, \bibinfo{person}{Xiao Huang}, \bibinfo{person}{Yue Zhao}, \bibinfo{person}{Yushun Dong}, {and} \bibinfo{person}{Jundong Li}.} \bibinfo{year}{2022}\natexlab{}.
\newblock \showarticletitle{Contrastive attributed network anomaly detection with data augmentation}. In \bibinfo{booktitle}{\emph{PAKDD}}. \bibinfo{pages}{444--457}.
\newblock


\bibitem[Yang et~al\mbox{.}(2022b)]%
        {yang2022reinforcement}
\bibfield{author}{\bibinfo{person}{Ruichao Yang}, \bibinfo{person}{Xiting Wang}, \bibinfo{person}{Yiqiao Jin}, \bibinfo{person}{Chaozhuo Li}, \bibinfo{person}{Jianxun Lian}, {and} \bibinfo{person}{Xing Xie}.} \bibinfo{year}{2022}\natexlab{b}.
\newblock \showarticletitle{Reinforcement subgraph reasoning for fake news detection}. In \bibinfo{booktitle}{\emph{ACM KDD}}. \bibinfo{pages}{2253--2262}.
\newblock


\bibitem[Yang et~al\mbox{.}(2022a)]%
        {YangHXL22}
\bibfield{author}{\bibinfo{person}{Yuhao Yang}, \bibinfo{person}{Chao Huang}, \bibinfo{person}{Lianghao Xia}, {and} \bibinfo{person}{Chenliang Li}.} \bibinfo{year}{2022}\natexlab{a}.
\newblock \showarticletitle{Knowledge Graph Contrastive Learning for Recommendation}. In \bibinfo{booktitle}{\emph{SIGIR}}. \bibinfo{pages}{1434--1443}.
\newblock


\bibitem[Yoo et~al\mbox{.}(2023)]%
        {YooLSK23}
\bibfield{author}{\bibinfo{person}{Hyunsik Yoo}, \bibinfo{person}{Yeon{-}Chang Lee}, \bibinfo{person}{Kijung Shin}, {and} \bibinfo{person}{Sang{-}Wook Kim}.} \bibinfo{year}{2023}\natexlab{}.
\newblock \showarticletitle{Disentangling Degree-related Biases and Interest for Out-of-Distribution Generalized Directed Network Embedding}. In \bibinfo{booktitle}{\emph{{ACM} Web Conference ({WWW})}}. \bibinfo{pages}{231--239}.
\newblock


\bibitem[Zeng et~al\mbox{.}(2021)]%
        {fair_hin}
\bibfield{author}{\bibinfo{person}{Ziqian Zeng}, \bibinfo{person}{Rashidul Islam}, \bibinfo{person}{Kamrun~Naher Keya}, \bibinfo{person}{James Foulds}, \bibinfo{person}{Yangqiu Song}, {and} \bibinfo{person}{Shimei Pan}.} \bibinfo{year}{2021}\natexlab{}.
\newblock \showarticletitle{Fair {{Representation Learning}} for {{Heterogeneous Information Networks}}}.
\newblock \bibinfo{journal}{\emph{ICWSM}}  \bibinfo{volume}{15} (\bibinfo{year}{2021}), \bibinfo{pages}{877--887}.
\newblock


\bibitem[Zhao et~al\mbox{.}(2024)]%
        {zhao2023competeai}
\bibfield{author}{\bibinfo{person}{Qinlin Zhao}, \bibinfo{person}{Jindong Wang}, \bibinfo{person}{Yixuan Zhang}, \bibinfo{person}{Yiqiao Jin}, \bibinfo{person}{Kaijie Zhu}, \bibinfo{person}{Hao Chen}, {and} \bibinfo{person}{Xing Xie}.} \bibinfo{year}{2024}\natexlab{}.
\newblock \showarticletitle{Competeai: Understanding the competition behaviors in large language model-based agents}. In \bibinfo{booktitle}{\emph{ICML}}.
\newblock


\bibitem[Zhu et~al\mbox{.}(2021)]%
        {zhu2021deep}
\bibfield{author}{\bibinfo{person}{Yanqiao Zhu}, \bibinfo{person}{Weizhi Xu}, \bibinfo{person}{Jinghao Zhang}, \bibinfo{person}{Qiang Liu}, \bibinfo{person}{Shu Wu}, {and} \bibinfo{person}{Liang Wang}.} \bibinfo{year}{2021}\natexlab{}.
\newblock \showarticletitle{Deep graph structure learning for robust representations: A survey}.
\newblock \bibinfo{journal}{\emph{arXiv:2103.03036}} (\bibinfo{year}{2021}).
\newblock


\bibitem[Zimek et~al\mbox{.}(2012)]%
        {zimekSurveyUnsupervisedOutlier2012}
\bibfield{author}{\bibinfo{person}{Arthur Zimek}, \bibinfo{person}{Erich Schubert}, {and} \bibinfo{person}{Hans-Peter Kriegel}.} \bibinfo{year}{2012}\natexlab{}.
\newblock \showarticletitle{A Survey on Unsupervised Outlier Detection in High-Dimensional Numerical Data}.
\newblock \bibinfo{journal}{\emph{Statistical Analysis and Data Mining}} \bibinfo{volume}{5}, \bibinfo{number}{5} (\bibinfo{year}{2012}), \bibinfo{pages}{363--387}.
\newblock


\end{thebibliography}
\bibliographystyle{ACM-Reference-Format}

\end{document}